\documentclass[a4paper]{article}

\usepackage{amsmath}
\usepackage{amssymb}
\usepackage{hyperref}
\usepackage{subcaption}
\usepackage{graphicx}
\usepackage{natbib}
\usepackage{siunitx}
\usepackage{mathtools}
\usepackage{algorithm2e}
\usepackage{multirow}

\providecommand{\keywords}[1]
{
  \small	
  \textbf{\textit{Keywords---}} #1
}

\newtheorem{property}{Property}
\usepackage{authblk}
  \title{A new data structure for efficient
  search on isovists}

\author[1]{Florent Gr\'{e}lard} 

\author[1]{Mehdi Ayadi}  

\author[1]{Mihaela Scuturici} 

\author[1]{Serge Miguet}

 \affil[1]{Universit\'{e} de Lyon, Lyon 2
   LIRIS F-69676 Lyon, France}
 
\affil[ ]{{\{florent.grelard, mehdi.ayadi, mihaela.scuturici, serge.miguet\}@univ-lyon2.fr}}

\begin{document}
\maketitle

\begin{abstract}
  Spatial data structures allow to make efficient queries on
  Geographical Information Systems (GIS). Spatial queries involve the
  geometry of the data, such as points, lines, or polygons. For
  instance, a spatial query could poll for the nearest restaurants
  from a given location. Spatial queries can be solved exhaustively by
  going through the entire data, which is prohibitive as the number of
  data points increases. In this article, we are interested in making
  efficient queries on infinitely long geometrical shapes. For
  instance, angular sectors, defined as the intersection of two
  half-spaces, are infinitely long. However, regular spatial data
  structures are not adapted to these geometrical shapes. We propose a
  new method allowing to make efficient spatial queries on angular
  sectors (i.e. whether a point is inside an angular sector). It
  builds a R-tree from the dual space of angular sectors. An extensive
  evaluation shows our method is faster than using a regular R-tree.
\end{abstract}

\keywords{Spatial data structure, R-tree, isovist, duality, projective
  geometry}

\section{Introduction}
\label{sec:intro}
\begin{figure}[ht]
  \centering
  \begin{subfigure}[t]{0.45\linewidth}
    \centering \includegraphics[width=0.75\textwidth]{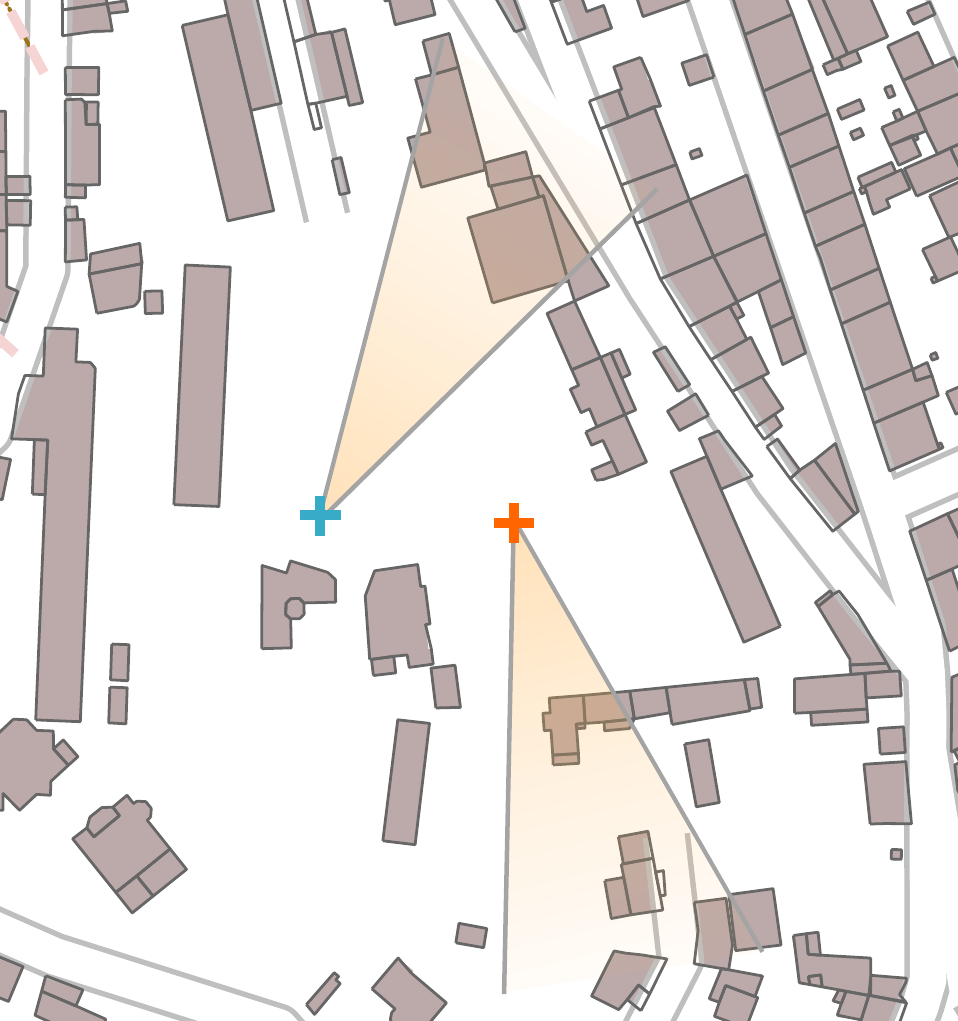}
    \caption{}
    \label{subfig:urban_cones}
  \end{subfigure}%
  \begin{subfigure}[t]{0.45\linewidth}
    \centering
    \includegraphics[width=0.75\textwidth]{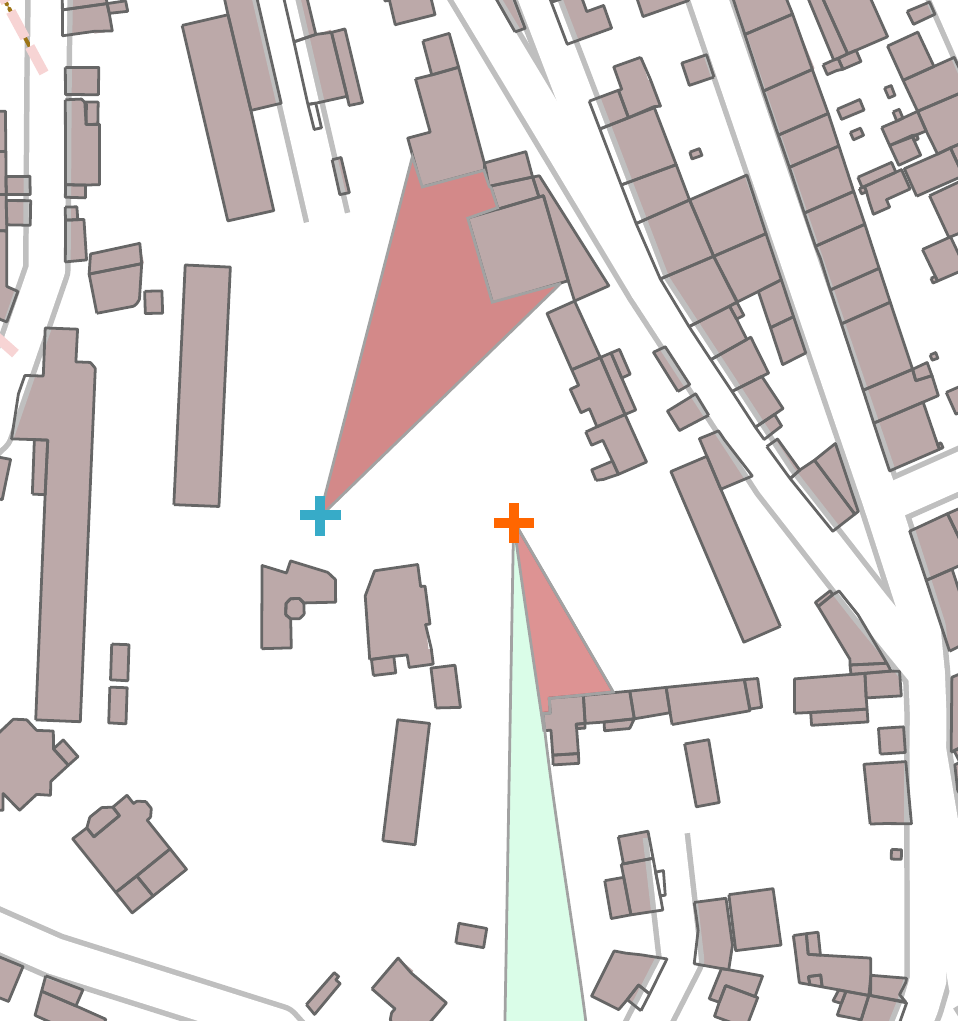}
    \caption{}
    \label{subfig:urban_isovist}
  \end{subfigure}%
  \caption{Typical spatial configuration in an urban environment with
    buildings (in gray) and visible space (in white). Two photographs (in
    blue and orange) with their associated (a) field of view (in pale
    orange) and (b) isovist (in red and green). (b) The isovist
    corresponds to the visible area from the point of view. The
    presence or absence of buildings defines areas with blocked
    visibility (in red) or free visibility (in green).}
  \label{fig:isovist}
\end{figure}

The integration of the visible space from the point of view of
pedestrians is an important step in urban planning~\cite{Turner01}.
The visible space can be extracted from a photographic investigation,
which is a documentary practice led by professional photographers
(e.g. \citep{Lawson88}). In a city, photographic investigations result
in a collection of photographs, from which hypotheses are made about
the inhabitants' lifestyle, the architectural process, or how the
visible space is perceived in the city. Usually, these hypotheses are
based on the fact that a small number of photographs share certain
traits, such as architectural features. The process of highlighting
these traits is impeded as the number of photographs increases. One
approach is to group photographs near a given location on a map based
on the assumption that geographically close photographs share
similarities. However, the content of geographically close photographs
can be completely different, in particular in cities, where the
landscape is rapidly changing. Another approach is to group
photographs based on the points they visualize, that is to say their
fields of view, as described by \cite{Ay08} or \cite{Lu14}. This can
also be achieved by searching the photographs based on their isovists.


The isovist is defined as the area of visible space from a given point
of view \citep{Benedikt79}. In an urban landscape, it allows to
analyze the space layout by characterizing the impact of a building on
the field of view of the photograph (see Fig.~\ref{fig:isovist}). It
is computed from the position and direction of the photograph. For
instance, the authors in \cite{Benedikt79} propose an algorithm to
compute the isovist in 2D by joining the intersections between the
building segments and rays drawn from the current position, in all
directions. The drawback of such a method is that it is dependent on
the number of rays: if it is too low, the resulting isovist is not
accurate, and if it is too high, the computation time is prohibitive.
The method introduced in \cite{Suleiman13} maps building segments to
the angle values they occupy in the field of view. The building
segments are processed based on their distance: the closest segments
to the photograph's location are considered first. After they are
mapped to an angle value, they are not considered further in the
analysis.

\begin{figure}[ht]
  \centering
  \begin{subfigure}[t]{0.49\linewidth}
    \centering
    \includegraphics[width=0.8\linewidth]{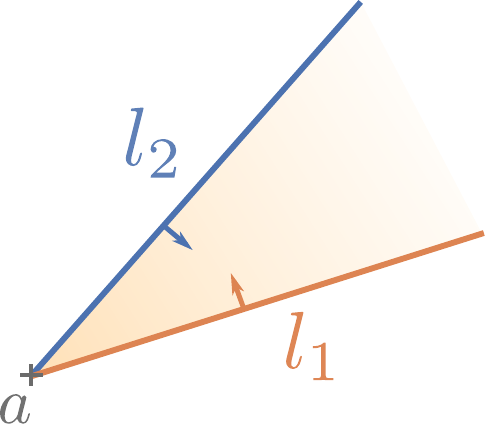}
    \caption{}
    \label{subfig:angularsector1}
  \end{subfigure}%
  \begin{subfigure}[t]{0.49\linewidth}
    \centering
    \includegraphics[width=0.9\linewidth]{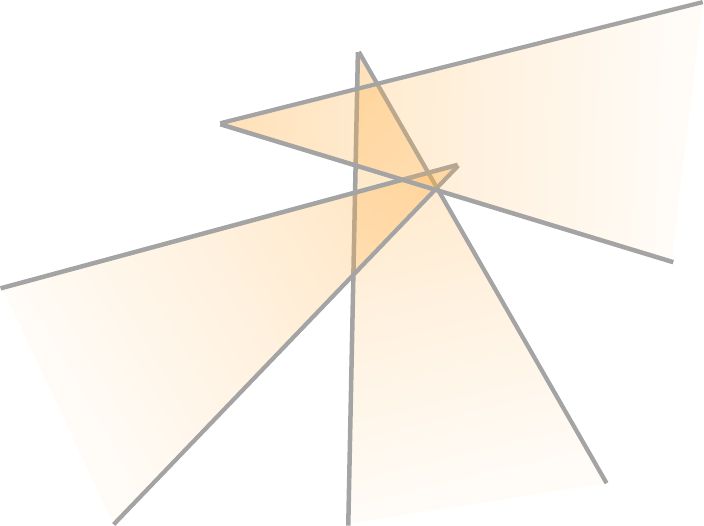}
    \caption{}
    \label{subfig:angularsector2}
  \end{subfigure}%
  \caption{(a) Angular sector: intersection between two half-spaces:
    above the lower line $l_1$ (in orange) and below the upper line
    $l_2$ (in blue). The intersection between the two lines
    corresponds to the apex $a$ of the sector. (b) Several angular
    sectors with different position and direction.}
  \label{fig:angularsector}
\end{figure}

Regardless of the method used for the isovist computation, the
visibility is \textbf{blocked} when buildings obstruct the field of
view (see Fig.~\ref{subfig:urban_isovist}). In this case, the isovist
is a polygon, that is to say, it has a bounded geometrical shape. On
the contrary, when there is no building, the visibility is
\textbf{free}, and the geometrical shape of the isovist is infinitely
long. In this article, we focus on these infinitely long shapes in 2D.

In the following, the free-visibility areas in isovists are called
angular sectors. Let $l_1: y = a_1x + b_1$ and $l_2: y = a_2x + b_2$
be non-parallel lines which intersect at $a$, with
$\tan^{-1}(a_1) < \tan^{-1}(a_2)$. An angular sector of apex $a$ is
defined as the intersection between the half-spaces above $l_1$ and
below $l_2$ (see Fig.~\ref{fig:angularsector}). This point is called
the apex $a$ of the angular sector. An angular sector can be also represented thanks
to three variables: the apex, the direction of the bisector, and the
angle value at the apex.

The query ``find all photographs which visualize a given point'' can
be achieved in logarithmic time with the R-tree \citep{Guttman84}, for
photographs with a bounded and small isovist. This is not the case for
photographs with infinitely long geometrical shapes, i.e. angular
sectors. In this article, we are interested in solving the following spatial
query efficiently: ``find all angular sectors which contain a given
point''. This is achieved by computing the intersection between the
point and all the angular sectors. Such a query can be solved with an
exhaustive search of all angular sectors, which is prohibitive as the
number of photographs increases. Another approach would be to
partition the space with evenly distributed angle intervals, from 0 to
\ang{360}, and to return the angular sectors present in the selected
interval. However, this is not adapted to angular sectors with
different positions because they can intersect despite having
different angle values (see Fig.~\ref{subfig:angularsector2}).
Moreover, logarithmic time could not be achieved, as it is not a
recursive partition of the
space. 

In this article, we propose a new spatial data structure, called dual
R-tree, to search efficiently through 2D angular sectors. The
contributions of this paper are listed below.

\begin{itemize}
\item We review related methods from the literature in detail, and
  explain why they are not suited to the problem of searching
  efficiently through angular sectors.
\item We describe two dual transforms which allow to convert
  infinitely long geometrical shapes to finite coordinates. These
  transforms are based on the existing duality between points and
  lines in the Euclidean space. This allows to convert a line to
  finite coordinates. We explain how the transforms are applicable to
  angular sectors.
\item We introduce a new spatial data structure, called the dual
  R-tree. To the best of our knowledge, this is the first work on
  indexing angular sectors and infinitely-long geometrical shapes. The
  dual R-tree is an R-tree built on the dual coordinates of angular
  sectors. We also explain how to handle the limit-cases of the dual
  transforms in order to obtain a proper set of finite coordinates.
\item Our method is evaluated on both synthetically generated data,
  and real-world data. The results show our method provides better
  optimization of the space inside of the tree, which translates in
  faster search times than the baseline methods. On the synthetic
  dataset, our method is at least five times faster than the baseline
  methods. These results are confirmed on the real-world dataset.
\end{itemize}

The article is organized as follows. In
Section~\ref{sec:previousworks}, we review the existing accelerating
data structures. In Section~\ref{sec:dual}, we recall the basics of
dual transforms, which allow us to transform infinitely long
geometrical shapes to bounded shapes. Section~\ref{sec:proposedmethod}
describes our spatial data structure for angular sectors, based on
dual transforms. In Section~\ref{sec:evaluation}, we evaluate our
method by comparison to a regular R-tree.

\section{Related work}
\label{sec:previousworks}

\begin{figure}[ht]
  \centering \includegraphics[width=0.35\linewidth]{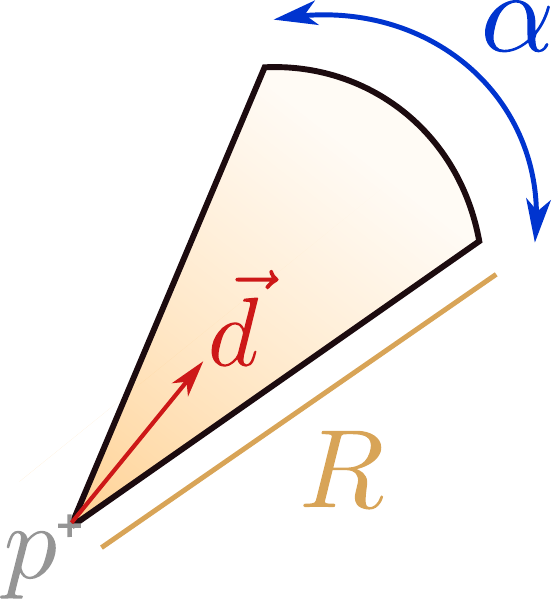}
  \caption{Field of view, defined by four variables \citep{Ay08}: the
    position $p$, the angle value $\alpha$, the direction vector
    $\vec{d}$ and the radius $R$.}
  \label{fig:fov}
\end{figure}

The authors in~\cite{Ay08} fully characterize a \textbf{field of view}
of a photograph by four variables: the position (latitude and
longitude values), the angle value, the direction and the visible
distance (see Fig.~\ref{fig:fov}). The latter variable corresponds to
the restricting radius of the field of view and bounds the geometrical
shape. This means the field of view is the bounded counterpart of
angular sectors. Since, to the best of our knowledge, angular sectors
have not been considered in the literature, this review is focused on
existing methods on fields of views. Several accelerating spatial data
structures were introduced in the literature for this type of data.
More generally, the existing methods can be classified into two
categories: the \textit{space-driven} structures, and the
\textit{data-driven} structures.

The \textit{space-driven} methods recursively partition the space,
which allows to discard large portions of the space in the search
procedure. They are comprised, among many others of the k-d tree
\citep{Bentley75}, the quadtree~\citep{Finkel74}, Voronoi diagrams
\citep{Wiley00}, and the grid \citep{Nievergelt84}. The grid
partitions the space into regular rectangles, which is adapted to
points, but not adapted to fields of view. The authors in \cite{Ma13}
propose a new grid structure for fields of view by defining three
different hierarchical levels of space subdivision. The first level
contains cells associated to the positions of the fields of view, the second level
corresponds to their radius, and the third level is associated to their
direction. During the search procedure, the positions and directions
are processed separately, with predefined radius search range. For
this reason, this structure is not adapted to geometrical shapes with
infinite radius, such as angular sectors.

\begin{figure}[ht]
  \centering
  \begin{subfigure}[t]{0.33\linewidth}
    \centering \includegraphics[width=0.9\linewidth]{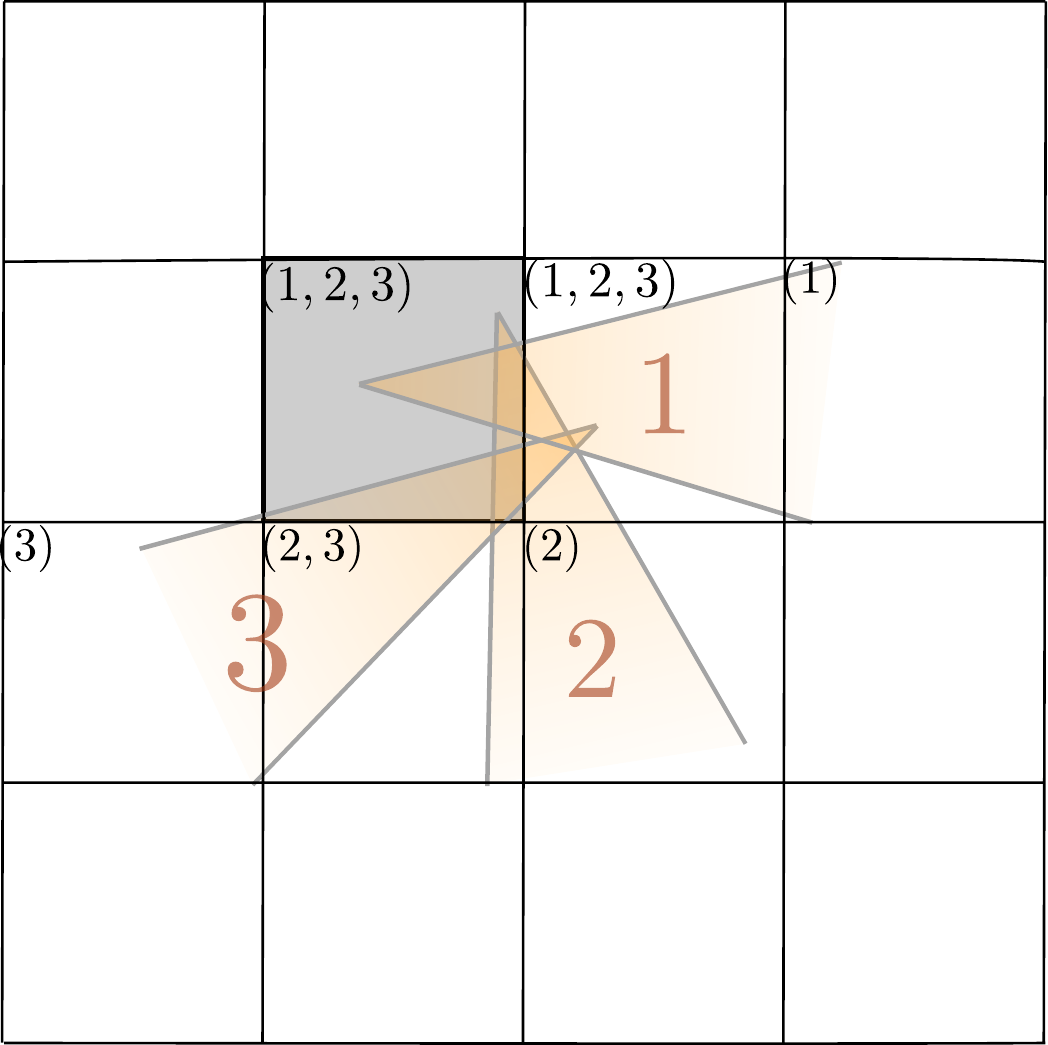}
    \caption{}
    \label{subfig:grid1}
  \end{subfigure}%
  \begin{subfigure}[t]{0.33\linewidth}
    \centering \includegraphics[width=0.9\linewidth]{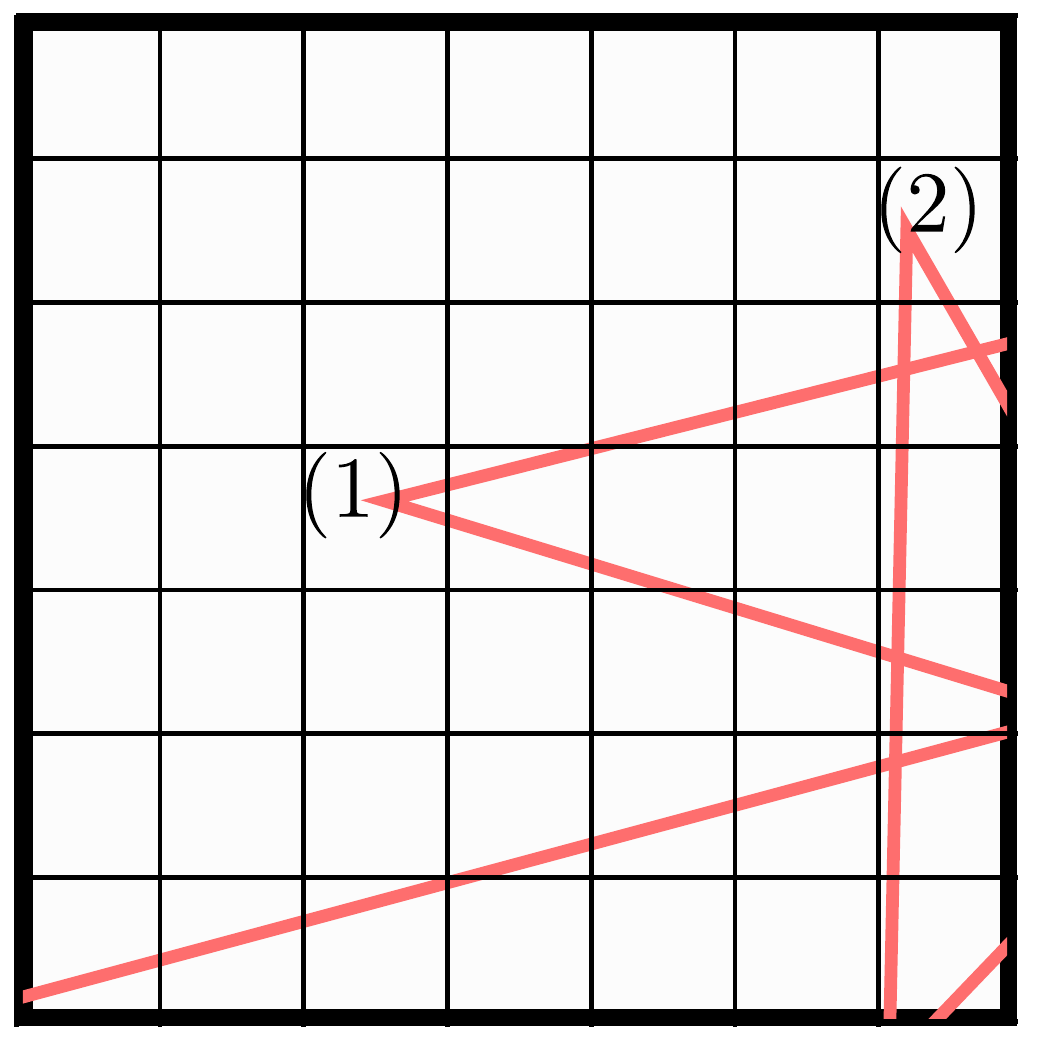}
    \caption{}
    \label{subfig:grid2}
  \end{subfigure}%
  \begin{subfigure}[t]{0.33\linewidth}
    \centering \includegraphics[width=0.95\linewidth]{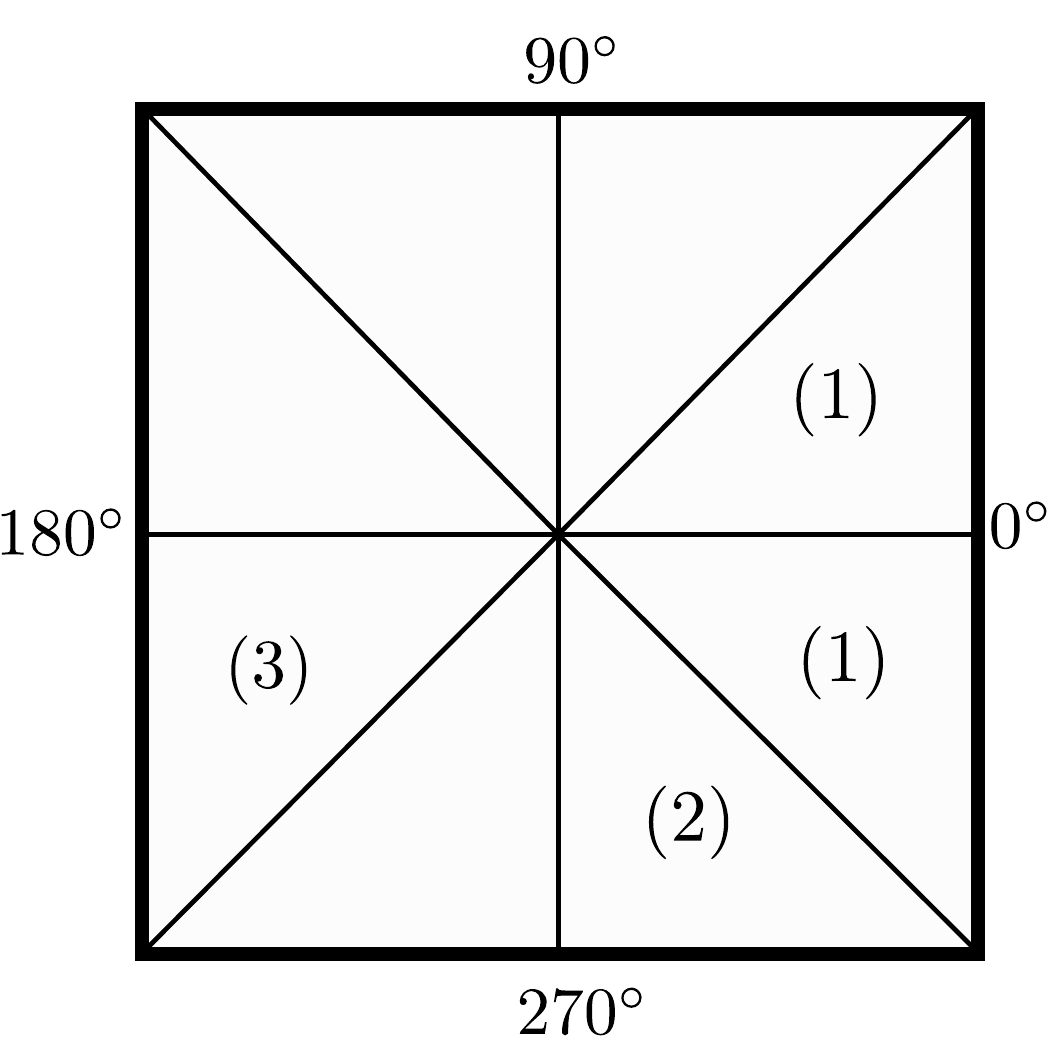}
    \caption{}
    \label{subfig:grid3}
  \end{subfigure}%
  \caption{Example of a space-driven structure: the grid for three
    fields of view (denoted from 1 to 3, in orange), with three
    hierarchical levels \cite{Ma13}. The cell indices are noted in
    brackets inside the relevant cells. (a) First level, where cells
    indices correspond to the fields of view it contains. (b) Second
    level, close-up of the gray cell of Fig.~(a). The cell indices
    correspond to the position of the field of view. Third level,
    where the grid is divided in sectors of regular intervals, where
    sector indices correspond to the direction of the field of view. }
  \label{fig:grid}
\end{figure}

\begin{figure}[ht]
  \centering
  \begin{subfigure}[t]{0.48\linewidth}
    \centering \includegraphics[width=0.8\linewidth]{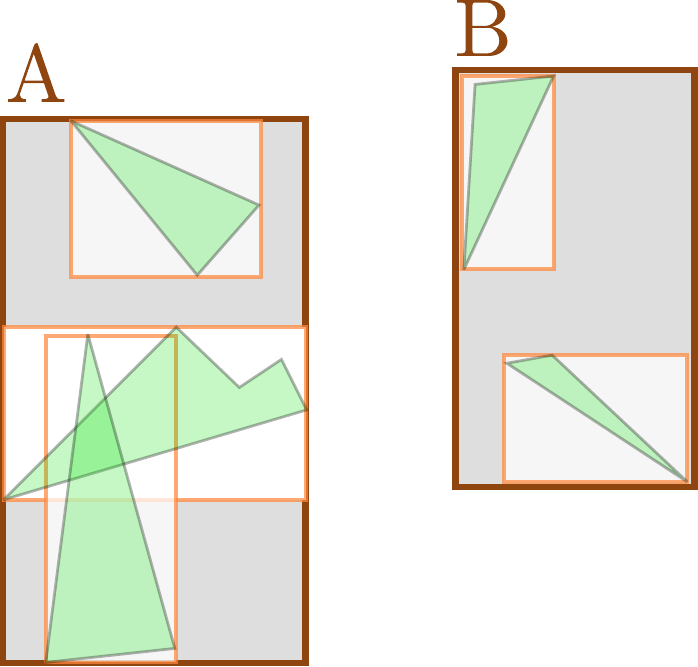}
    \caption{}
    \label{subfig:rtree_2}
  \end{subfigure}%
  \begin{subfigure}[t]{0.48\linewidth}
    \centering
    \includegraphics[width=0.87\linewidth]{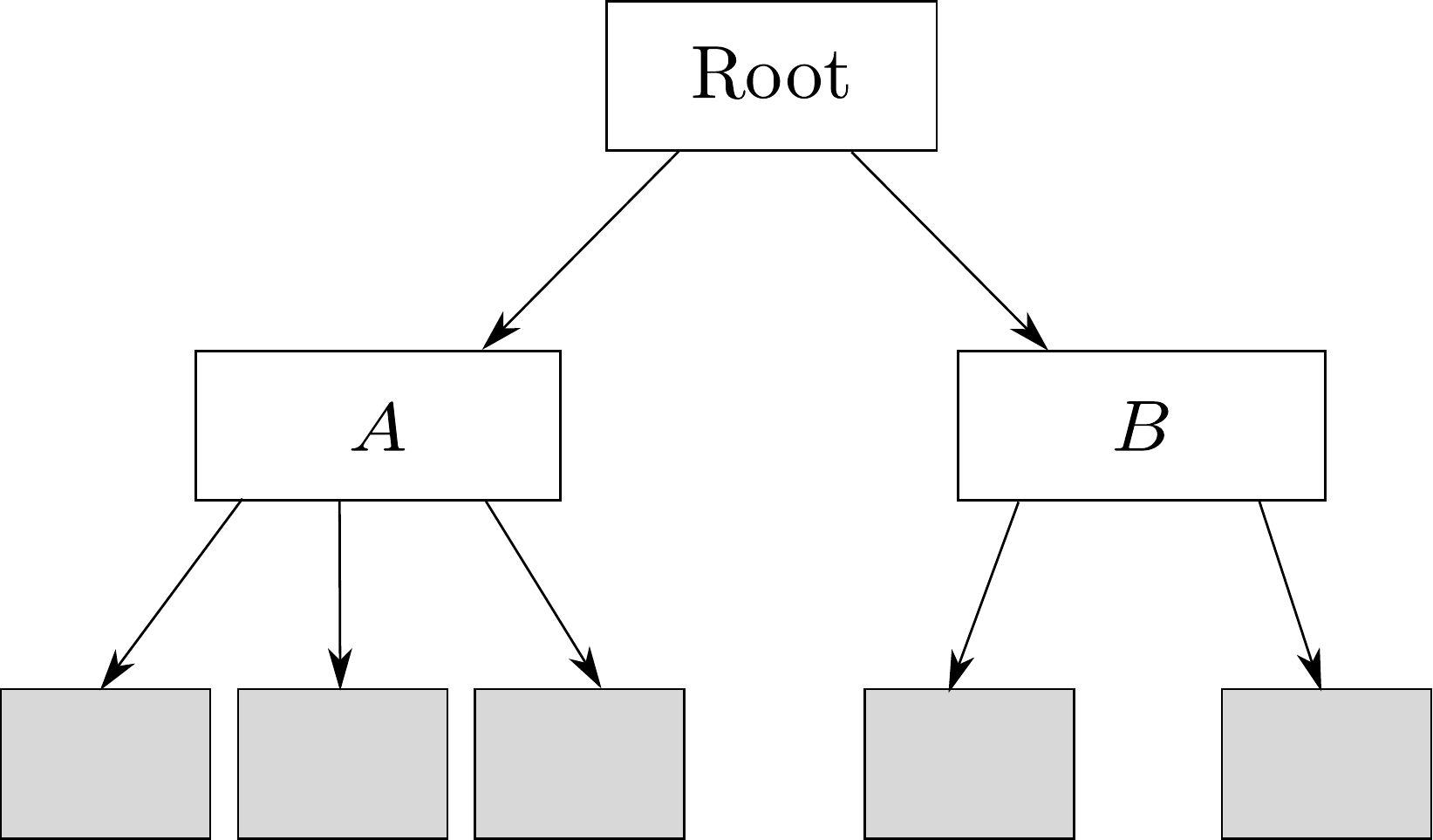}
    \caption{}
    \label{fig:rtree_tree}
  \end{subfigure}%
  \caption{Indexing in the R-tree: (a) geometrical shapes (polygons
    from fields of view, in green) are represented by their minimum
    bounding rectangles (in orange). The empty space covers a large
    area for each polygon (in white). Rectangles are grouped in
    enclosing rectangles (A and B). (b) Resulting R-tree from the
    polygons in Fig.~\ref{fig:rtree}, where the minimum bounding
    rectangles correspond to the leaves of the tree (in gray).}
  \label{fig:rtree}
\end{figure}

The \textit{data-driven} structures revolve around the geometrical
shape of the data. They are comprised, among many others of the
R-tree~\citep{Guttman84}. The R-tree is a hierarchical data structure
of rectangles, defined in~\cite{Guttman84}, and based on the
B+-tree~\citep{Comer79}. The R-tree is a height-balanced tree, where
the geometrical shapes are represented by their minimum bounding
rectangles~(see Fig.~\ref{fig:rtree}). These rectangles constitute the
leaves in the R-tree~(see Fig.~\ref{fig:rtree_tree}). The bounding
rectangles are then grouped in larger, enclosing rectangles. A query
which does not intersect an enclosing rectangle in the higher levels
of the hierarchy does not intersect any of the enclosed rectangles.
The branching factor determines the maximum number of children allowed
at each node.

Several algorithms have been proposed to group rectangles so that a
query yields as few false positive errors as possible. Usually, they
involve optimizing various criteria, such as the coverage and overlap.
These metrics quantify the amount of empty space inside a node of the
tree, that is to say the area of a minimum bounding rectangle which
does not contain the enclosed geometrical data. The \textbf{coverage}
is the ratio between the area covered by all enclosed rectangles and
the area of the enclosing rectangle. The \textbf{overlap} is the ratio
of the area covered by more than one rectangle with respect to the
area of the enclosing rectangle. The R-tree allows to achieve
$\mathcal{O}(\log n)$ query time under the assumption that both the
coverage and the overlap are minimized. When both the coverage
approaches one and the overlap is low, the amount of empty space
inside the enclosing rectangle is low. The authors in \citep{Sellis87}
defined the R+-tree, which aims at optimizing both the coverage and
the overlap, by splitting the minimum bounding rectangles and allowing
the geometrical shapes to be added in various leaves.

In \cite{Ay08}, the authors use the minimum bounding rectangle of the
field of view. The drawback of such a method is that the resulting
rectangles in the hierarchy have a large amount of empty space. The
authors in \cite{Lu14}, \cite{Lu16}, and \cite{Lu17} propose to
minimize the amount of empty space by taking the direction of the
field of view into account. The fields of views are enclosed in a
minimum bounding field of view, which is a field of view enclosing its
children. The search algorithm restricts the search to fields of view
whose radius is less than a given distance. This structure is not
adapted to angular sectors, since their radius is infinite.

More generally, the spatial data structures reviewed in this
section are not adapted to infinitely long geometrical shapes, such as
angular sectors. Indeed, they all consider bounded shapes, either
during the insertion or the search procedure. In the next section, we
propose a new method to make efficient queries on angular sectors.

\section{Dual transforms}
\label{sec:dual}
In the following, we review the affine and dual transforms, which
convert infinitely long geometrical shapes to finite coordinates.

In the following, we are interested in \textbf{bi-angular sectors},
defined as the union of an angular sector, and its symmetric
counterpart with respect to the apex (see
Fig.~\ref{subfig:affinedual_1}).

\subsection{Affine dual transform}
\label{subsec:affine}

The duality between points and lines stems from the observation that
two points in the Euclidean space define a \textbf{single} line, and
two lines intersect a \textbf{single} point \citep{Poncelet1865}. A
line $l : y = ax + b$ has two constants $a$ and $b$, which define
coordinates of a point $p(a, b)$ in the parametric space of lines.
Conversely, a point defines the coefficients of a line. The affine
dual transform $\delta_A$ maps a primal point to a dual line, and a
primal line to a dual point. Let $p = (x_p, y_p) \in \mathbb{R}^2$ be
a point, and $l$ be a non-vertical line. The affine dual transform is
defined by:

\begin{align}
  p(x_p, y_p)~&~\xmapsto{\delta_A}~p^\star: y = x_px - y_p \\
  l: y = ax + b~&~\xmapsto{\delta_A}~l^\star(a, -b)
\end{align}

This transform is involutory, that is to say $(p^\star)^\star = p$.
This transform has two additional interesting properties: incidence and order.
Incidence means a point $p$ is on a line $l$ if and only if the point
$l^\star$ is on the line $p^\star$. Order means a point $p$ is above
the line $l$ if and only if the point $l^\star$ is above the line
$p^\star$.

\begin{figure}[t]
  \centering
    \begin{subfigure}[t]{0.5\textwidth}
    \centering
    \includegraphics[width=0.9\linewidth]{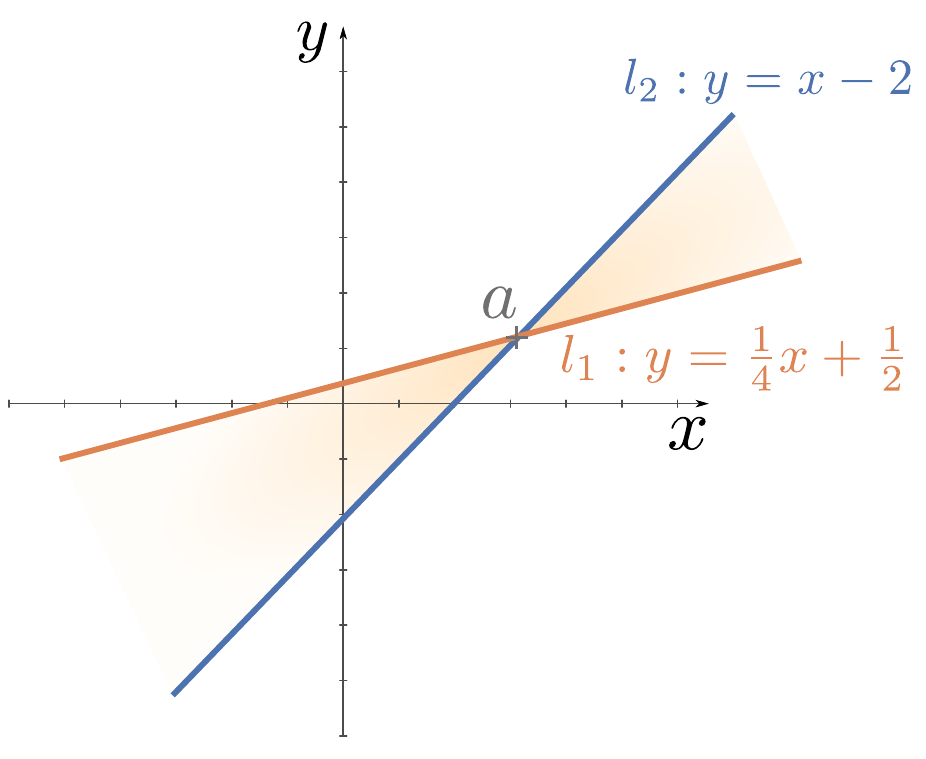}
    \caption{}
    \label{subfig:affinedual_1}
  \end{subfigure}%
  \begin{subfigure}[t]{0.5\textwidth}
    \centering
    \includegraphics[width=0.9\textwidth]{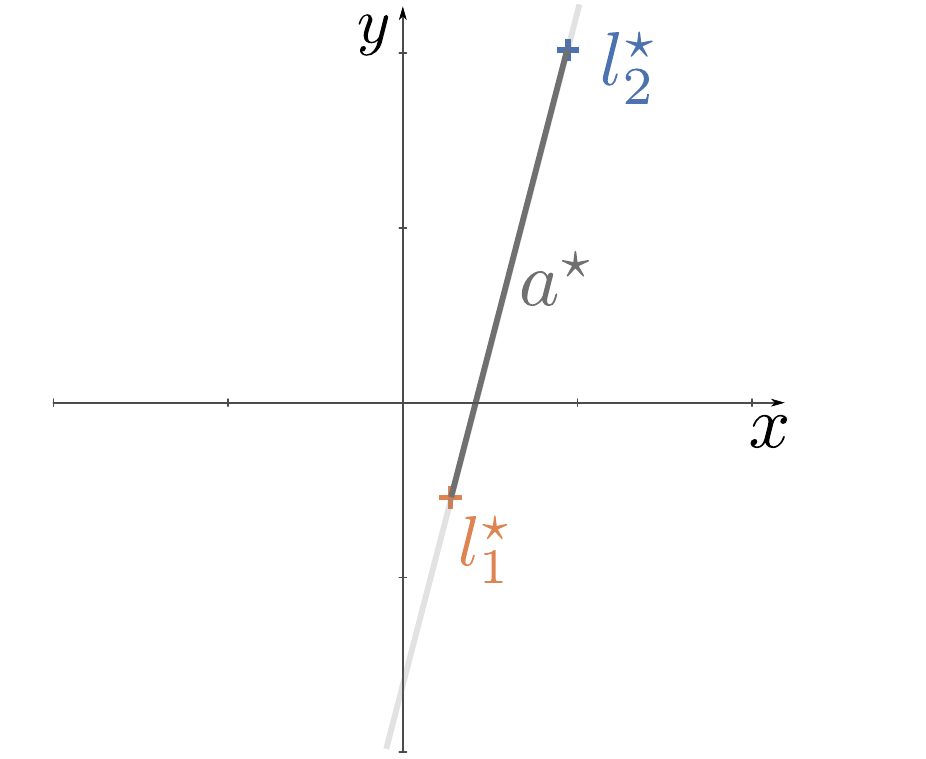}
    \caption{}
    \label{subfig:affinedual_2}
  \end{subfigure}%

  \caption{(a) Bi-angular sector of apex $a$ defined by two lines: lower
    and upper lines ($l_1$ and $l_2$ in blue and orange,
    respectively), (b) Affine dual of the angular sector from
    Fig.~\ref{subfig:affinedual_1}: segment joining the pair of dual
    coordinates ($l_1^\star$ and $l_2^\star$ in orange and blue). The
    apex $a^\star$ is a line (in light gray) joining these two
    coordinates.}
\end{figure}

The dual bi-angular sector of lower and upper lines $l_1$ and $l_2$ is
the segment joining the pair of coordinates $l_1^\star$ and $l_2^\star$ (see
Fig.~\ref{subfig:affinedual_2}). From the incidence property, it
follows that the apex of the angular sector corresponds to the line passing through the segment
$l_1 ^\star l_2 ^\star$ in the dual space. From the order property, it
follows that a point located inside an angular sector corresponds to a
dual line which crosses the segment $l_1 ^\star l_2 ^\star$.

However, this transform cannot handle vertical lines, since their
slope is not defined. Thus, two dual affine spaces are generally used:
one for the lines with slope values less than one (\textit{horizontal}
lines), and another for the lines with slope values greater than one
(\textit{vertical} lines). Vertical lines can be expressed in the form
of the equation $l: x = my + n$. Thus, the dual of a vertical line is
the point $l ^\star (m, -n)$.

\subsection{Polar dual transform}
The polar dual transform $\delta_P$, also called
$\rho-\theta$-parametrization, converts each line to a point and each
point to a sinusoid \citep{Radon1917}. It can be applied to vertical
lines, which makes it a strong candidate for angular sectors.

Let $l: y = ax+b$ be a line, and $H$ the orthogonal projection of the
origin onto $l$, then $\rho_H$ is the distance from the origin to $H$,
and $\theta_H$ is the the angle between the $x$-axis and the line
connecting the origin to $H$. The line $l$ can be rewritten in the
Hesse normal form as $l: \rho_H = x \cos \theta_H + y \sin \theta_H$.
Then, the polar dual transform $\delta_P$ is defined by:

\begin{align}
  p(x_p, y_p)~&~\xmapsto{\delta_P}~p^\star: \rho = x_p\cos \theta + y_p\sin \theta \\
  l: \rho_H = x \cos \theta_H + y \sin \theta_H~&~\xmapsto{\delta_P}~l^\star(\theta_H, \rho_H)
\end{align}

A point in the polar dual space is defined by a pair of coordinates
$(\theta, \rho)$, with $\theta$ ranging from $0$ to $\pi$ and
$\rho \in \mathbb{R}$, with $\rho < 0$ if the ordinate of the point in
the primal space is negative.

\begin{figure}[t]
  \centering
  \begin{subfigure}[t]{0.45\linewidth}
    \centering \includegraphics[width=0.8\linewidth]{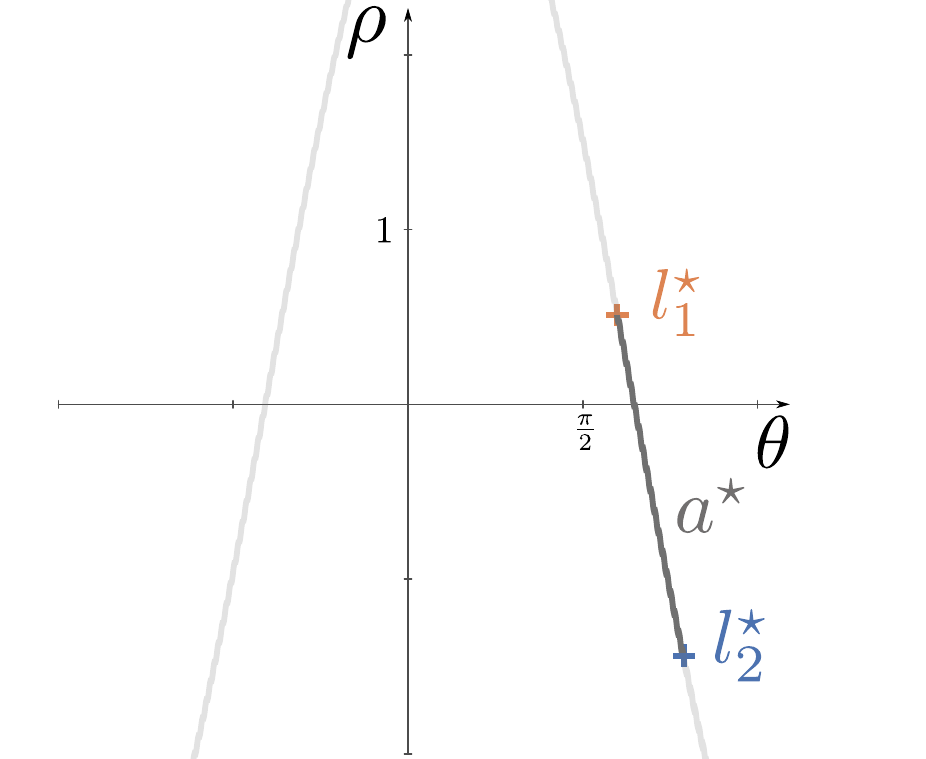}
    \caption{}
    \label{subfig:affinedual_3}
  \end{subfigure}%
  \begin{subfigure}[t]{0.45\linewidth}
    \centering \includegraphics[width=0.8\linewidth]{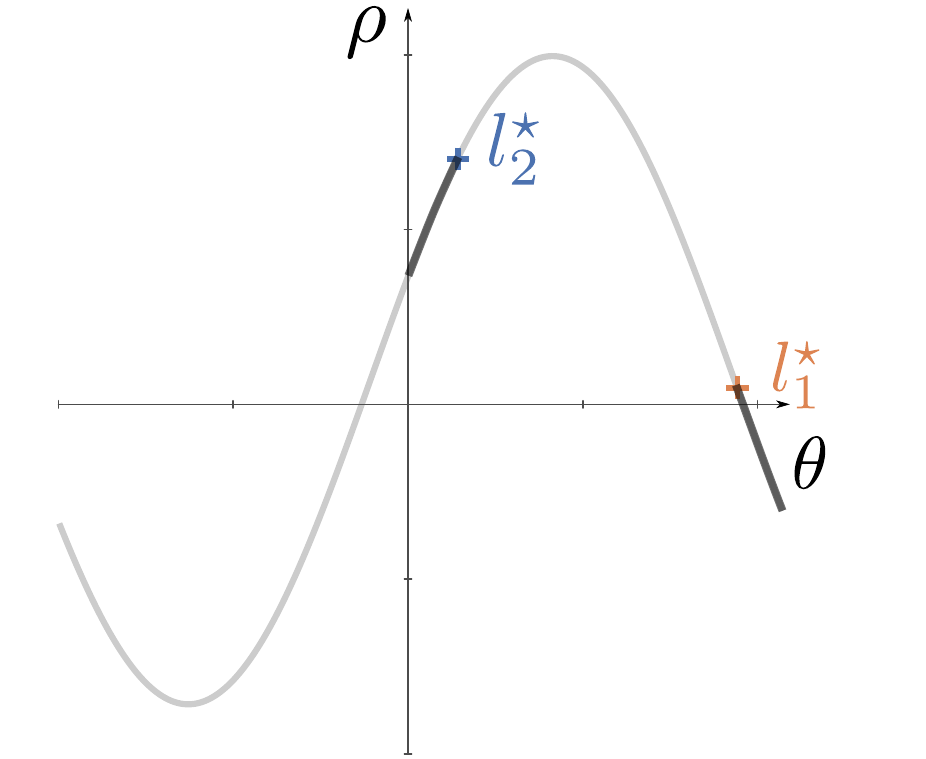}
    \caption{}
    \label{subfig:affinedual_4}
  \end{subfigure}%

  \caption{Polar dual of (a) the angular sector from
    Fig.~\ref{subfig:affinedual_1} and (b) from an angular sector
    containing a vertical line. The dual angular sector is (a) the arc
    of the sinusoid $a^\star$, or (b) the
    complementary of this arc on $[0, \pi]$ for sectors containing
    vertical lines. The apex
    $a^\star$ is the sinusoid represented in light gray.}
  \label{subfig:affinedual_3}
\end{figure}%

This transform verifies the incidence property: a point $p$ is on a
line $l$ if the point $l^\star$ is on the sinusoid $p^\star$. However,
the order property does not hold for angular sectors containing a
vertical line. Indeed, the lower line is transformed into a dual point
with a $\theta$ value close to $\pi$ and the upper line into a dual
point with a $\theta$ value close to 0.

This implies the following properties (see
Fig.~\ref{subfig:affinedual_3}):
\begin{property}
  For lines which \textbf{do not contain} a vertical line, a point $p$
  is located between two lines $l_1$ and $l_2$ if and only if the
  sinusoid $p^\star$ crosses the arc joining $l_1^\star$ and
  $l_2^\star$.
\end{property}

\begin{property}
  For lines which \textbf{contain} a vertical line, a point $p$ is
  located between two lines $l_1$ and $l_2$ if and only if the
  sinusoid $p^\star$ crosses the complementary of the arc joining
  $l_1^\star$ and $l_2^\star$, on the interval $[0, \pi]$.
\end{property}

The dual angular sector of lower and upper lines $l_1$ and $l_2$ is the
arc of sinusoid joining $l_1^\star$ and $l_2^\star$ for angular
sectors which do not contain a vertical line, and the complementary of
this arc on $[0, \pi]$ for sectors containing a vertical line. The
dual apex $a^\star$ is the sinusoid passing through this arc.

\section{Proposed method}
\label{sec:proposedmethod}

Our method builds a spatial data structure from a dual transform,
either the affine, or the polar dual. In the dual space, a bi-angular
sector corresponds to a pair of finite coordinates. Thus, these dual
coordinates can be inserted inside a hierarchical data structure with
minimized overlap and coverage.

Our method is not dependent on the spatial data structure. We use the
R-tree because it comes with various optimization strategies for split
and loading, and is balanced, as opposed to the k-d tree. In the
following, our method is referred to as the \textbf{dual R-tree}, and
we describe its insertion and search procedures.

\begin{figure*}[ht]
  \centering
  \begin{subfigure}[t]{0.33\textwidth}
    \centering \includegraphics[width=0.8\textwidth]{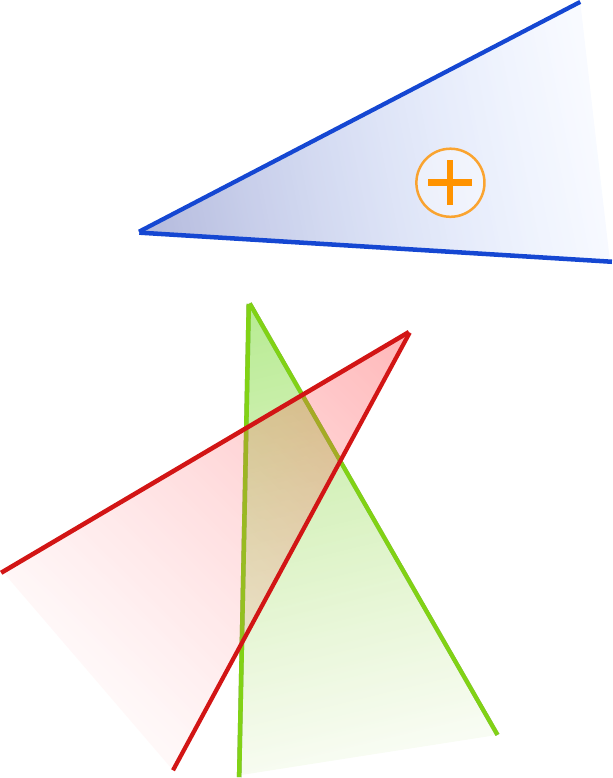}
    \caption{}
    \label{subfig:workflow_1}
  \end{subfigure}%
  \begin{subfigure}[t]{0.33\textwidth}
    \centering \includegraphics[width=0.87\textwidth]{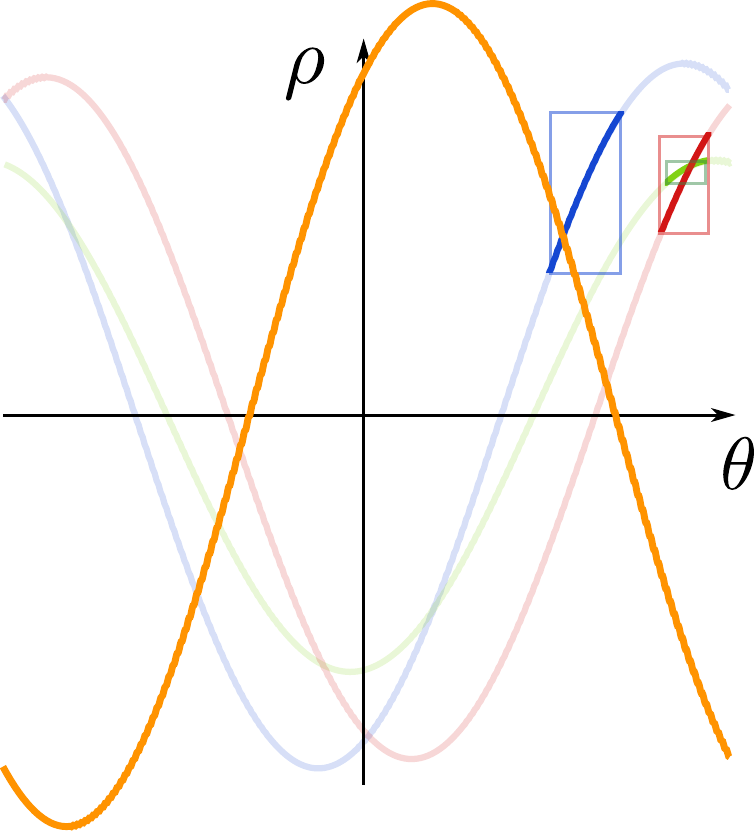}
    \caption{}
    \label{subfig:workflow_2}
  \end{subfigure}%
  \begin{subfigure}[t]{0.33\textwidth}
    \centering \includegraphics[width=0.95\textwidth]{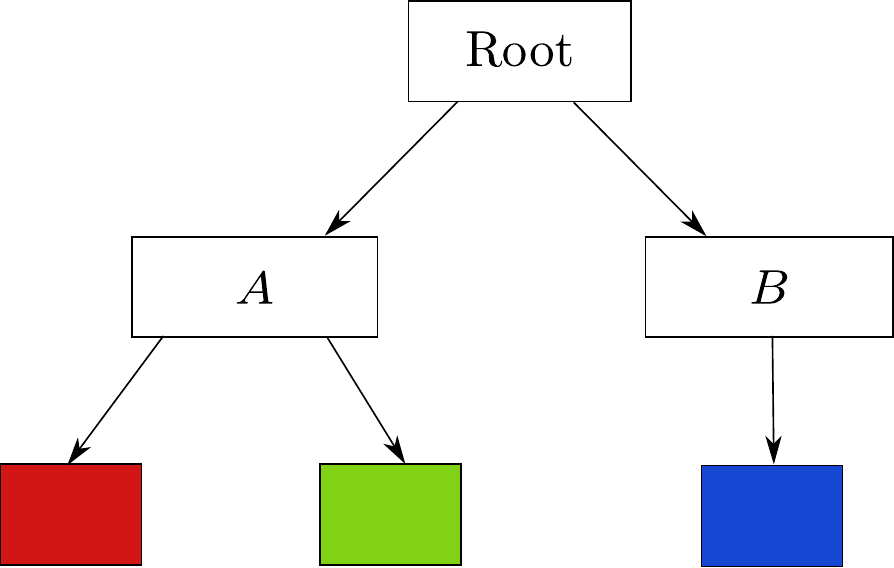}
    \caption{}
    \label{subfig:workflow_3}
  \end{subfigure}%
  \caption{Overview of our method for the insertion and search of
    angular sectors inside the spatial data structure. (a) Primal
    angular sectors (in red, green and blue). We want to find all
    sectors containing the query point (in orange). The red and green
    angular sectors intersect. (b) Transformation of the primal
    angular sectors of Fig.~(a) by the polar dual transform. The
    primal apices are transformed into sinusoids (in pale colors). The
    dual angular sectors correspond to arcs of sinusoids (in bright
    colors). The bounding rectangle of each arc of sinusoid form the
    leaves of our spatial data structure. The query point is
    transformed into a sinusoid (in orange) which intersects the
    rectangle of the blue dual angular sector. (c) Dual spatial data
    structure where each color corresponds to the respective bounding
    rectangle from Fig.~(b). }
  \label{fig:workflow}
\end{figure*}

Regarding the insertion procedure, each dual angular sector is
inserted as the minimum bounding rectangle of its pair of dual
coordinates (see Fig.~\ref{fig:workflow}). These bounding rectangles
correspond to the leaves inside the dual R-tree. The leaves also
contain a reference to the primal angular sector in order to
facilitate their retrieval. Strategies and optimizations from the
R+-tree (see Section~\ref{sec:previousworks}) are used to minimize the
coverage and the overlap in the higher levels of the hierarchy.

Regarding the search algorithm, we are interested in three possible
spatial queries: point, range and directional queries. A search by
point returns all the angular sectors containing this point, a search
by range returns the angular sectors intersecting a circular query
range, and search by direction returns the angular sectors
intersecting a line.

Regarding point queries, the primal query point is converted to its
dual, that is to say a line for the affine dual transform, or an arc
of a sinusoid for the polar dual transform. The primal point is inside
an angular sector if the dual of the query point intersects the
segment joining the dual coordinates. The nodes of the R-tree are
traversed by a breadth-first search, starting from the root of the
tree. If a node intersects a dual query point, its children are
considerer further in the search procedure, otherwise, its children
are discarded. During the search procedure, an enclosing rectangle is
discarded if it does not intersect the dual of the query point. This
method might yield some false positives, in particular if the dual of
the query point intersects the bounding rectangle but not the dual of
the apex $a^\star$. However, in our case, the angular sectors have a
small angle (typically lower than \ang{10}), so the difference in
$\theta$-coordinates for the pair of dual points is small, and the
amount of empty space in the rectangle is negligible compared to the
area of the enclosing rectangle. The assumptions we make on the angle
distribution of our data is discussed further in
Section~\ref{sec:evaluation}.

\IncMargin{1em} \RestyleAlgo{boxruled} \LinesNumbered
\begin{algorithm}[h]
  \KwData{$S$: primal angular sectors, $R$: empty R-tree}
  \KwResult{$R$: Dual R-tree} \ForAll{$s \in S$}{
    $l_1 \leftarrow s.lowerLine$\; $l_2 \leftarrow s.upperLine$\;
    $l_1^\star \leftarrow \delta(l_1)$\;
    $l_2^\star \leftarrow \delta(l_2)$\;
    $r \leftarrow \texttt{MinimumBoundingRectangle}(l_1^\star,
    l_2^\star)$\;
    $R.\texttt{insert}(\{\texttt{rectangle}: r, \texttt{pointer}:
    s\})$\; } \KwRet{R}
  \caption{Building the dual R-tree.}
  \label{algo:dualrtree}
\end{algorithm}
\DecMargin{1em}

A directional query is transformed into a single point in the dual
space, so it can be processed similarly to point queries. A range query results in a collection of lines in the affine
dual space, and a collection of sinusoids in the polar dual space.
Processing this collection entirely does not result in an efficient
search procedure. In the following, the spatial queries are restricted
to points. The same methods are also applicable to directional
queries.

The search procedure returns a list of bi-angular sectors. However, the initial
query is to find \textbf{mono}-angular sectors (see
Section~\ref{sec:intro}). This means the resulting list must be
trimmed of the irrelevant half of the sectors. Let $n$ be the initial
length of the list of sectors, and $k$ the length of the list obtained
after the search on the dual R-tree, this operation is achieved in
$\mathcal{O}(k)$ time, with $k \ll n$.

The dual R-tree can be built from either the affine, or the polar dual
transforms. In the following, we describe the specificities pertaining
to the type of duality, regarding the insertion and search procedures.

\subsection{Affine dual R-tree}
The affine dual transform is not adapted to vertical lines, as
described in Section~\ref{subsec:affine}. It is necessary to maintain
two data structures: two R-trees $R_H$ and $R_V$ are built from the
two affine dual transforms. $R_H$ is associated with the transform
$T_H$ where the lines are expressed as $y = ax + b$ (for
\emph{horizontal} lines), and $R_V$ is associated with the transform
$T_V$ where the lines are expressed as $x = my + n$ (for
\emph{vertical} lines). The latter expression effectively rotates the
coordinate system ninety degrees clockwise. This means the initial
vertical lines appear horizontal in this coordinate system. Two dual
angular sectors $A_H^\star$ and $A_V^\star$ are obtained by the
transforms $T_H$ and $T_V$, respectively. In order to ensure the
overlap is minimized, only the dual angular sector with the lowest distance
between its pair of coordinates is kept, and its minimum bounding
rectangle is added to the corresponding R-tree. 

Both the R-trees $R_H$ and $R_V$ are polled during the search procedure.
Let $p=(x_p,y_p)$ be the input query point in the primal space, this
point is converted to two dual lines $p_H^\star: y = x_p x - y_p$ and
$p_V^\star: x = y_py - x_p$ by $T_H$ and $T_V$ respectively. The tree
$R_H$ is traversed by a breadth-first search, and we check whether the
rectangles intersect the line $p_H^\star$. The same procedure is repeated between the
rectangles of $R_V$ and the line $p_V^\star$.

\subsection{Polar dual R-tree}
\begin{figure}[ht]
  \centering \includegraphics[width=0.5\linewidth]{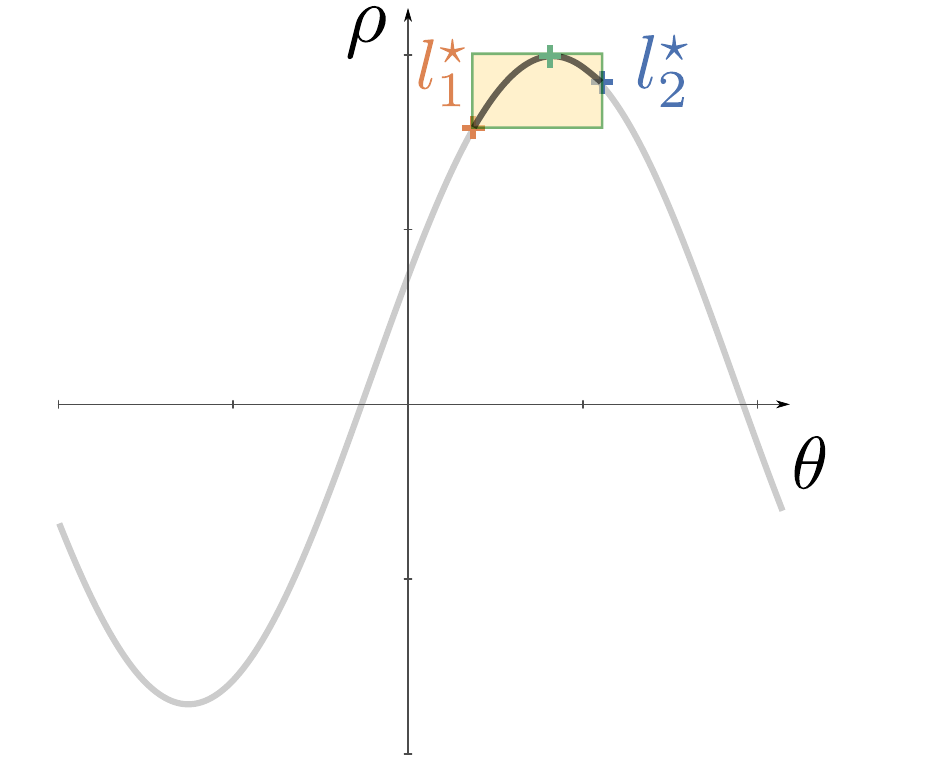}
  \caption{Corrected minimum bounding rectangle (in pale yellow) for a
    dual angular sector (coordinates in blue and orange) whose arc
    reaches a local extremum (in green).}
  \label{fig:extremum}
\end{figure}

The polar dual transform can handle vertical lines, which means only
one data structure needs to be maintained. For most angular sectors,
the enclosing rectangle is the bounding box of the dual of the angular
sector. However, two special cases need to be considered. First, for
an angular sector containing a vertical line, the dual transform does
not preserve the order (see Section~\ref{sec:dual}). In this case, two
bounding rectangles are inserted inside the R-tree, for each primal
lines delimiting the angular sector. The bounding rectangle for the
lower line spans the range $[\theta_{\text{lower}}, \pi]$, and the
bounding rectangle for the upper line spans the range
$[0, \theta_{\text{upper}}]$. Secondly, in the case where the arc of
the sinusoid joining two dual points reaches an extremum, the
enclosing rectangle of the dual angular sector is not large enough to
fully contain the arc of the sinusoid. In this case, the extremum is
computed and the enclosing rectangle is extended in order to contain
it (see Fig.~\ref{fig:extremum}).

Regarding the search procedure, the input query point $p=(a,b)$ is
transformed into a sinusoid in the dual space. The primal point is
inside a given angular sector if the sinusoid intersects the bounding
rectangle of the dual angular sector. In the following, we describe an
efficient algorithm which allows to check whether a sinusoid
intersects a rectangle.

Let $A = \sqrt{a^2 + b^2}$ and $\alpha = \tan^{-1}(\frac{b}{a})$, the
sinusoid $\rho = a \cos \theta + b \sin \theta$ can be rewritten as :
\[
  \rho = A \cos (\theta - \alpha)
\]

This form allows to compute the images and inverse images through the
sinusoid for the lower and upper bounds of a given enclosing
rectangle. If any of the images or inverse images are within the
bounds of an enclosing rectangle, then the sinusoid intersects this
rectangle (see Fig.~\ref{subfig:workflow_2}).

There are usually several inverse images through a given sinusoid. One
inverse image of $\rho$ is given by :
\[
  \theta(\rho) = \cos^{-1} (\frac{\rho}{A}) + \alpha
\].

Since $\alpha$ ranges from $-\frac{\pi}{2}$ to $\frac{\pi}{2}$,
$\theta(\rho)$ might not be in the range $[0, \pi]$. However, it is
straightforward to show that the period of the query sinusoid is
$2\pi$, so the other considered inverse image is :
\[
  \theta(\rho) = \cos^{-1} (\frac{\rho}{A}) + \alpha + 2\pi
\]

In total, two images and four inverse images are computed, for the
lower and upper bounds of the rectangle. The computation of the
intersection between a sinusoid and a rectangle is thus achieved in
$\mathcal{O}(1)$ time.

\section{Experimental evaluation}
\label{sec:evaluation}
In this section, we evaluate our dual spatial data structure. First,
we explain the experimental datasets and methods used for the
evaluation, and describe and analyze the results.

\subsection{Experimental setting}
Our data structure has been implemented using the RBush
library\footnote{\url{https://github.com/mourner/rbush}}.

\begin{figure}[ht]
  \centering \includegraphics[width=0.7\linewidth]{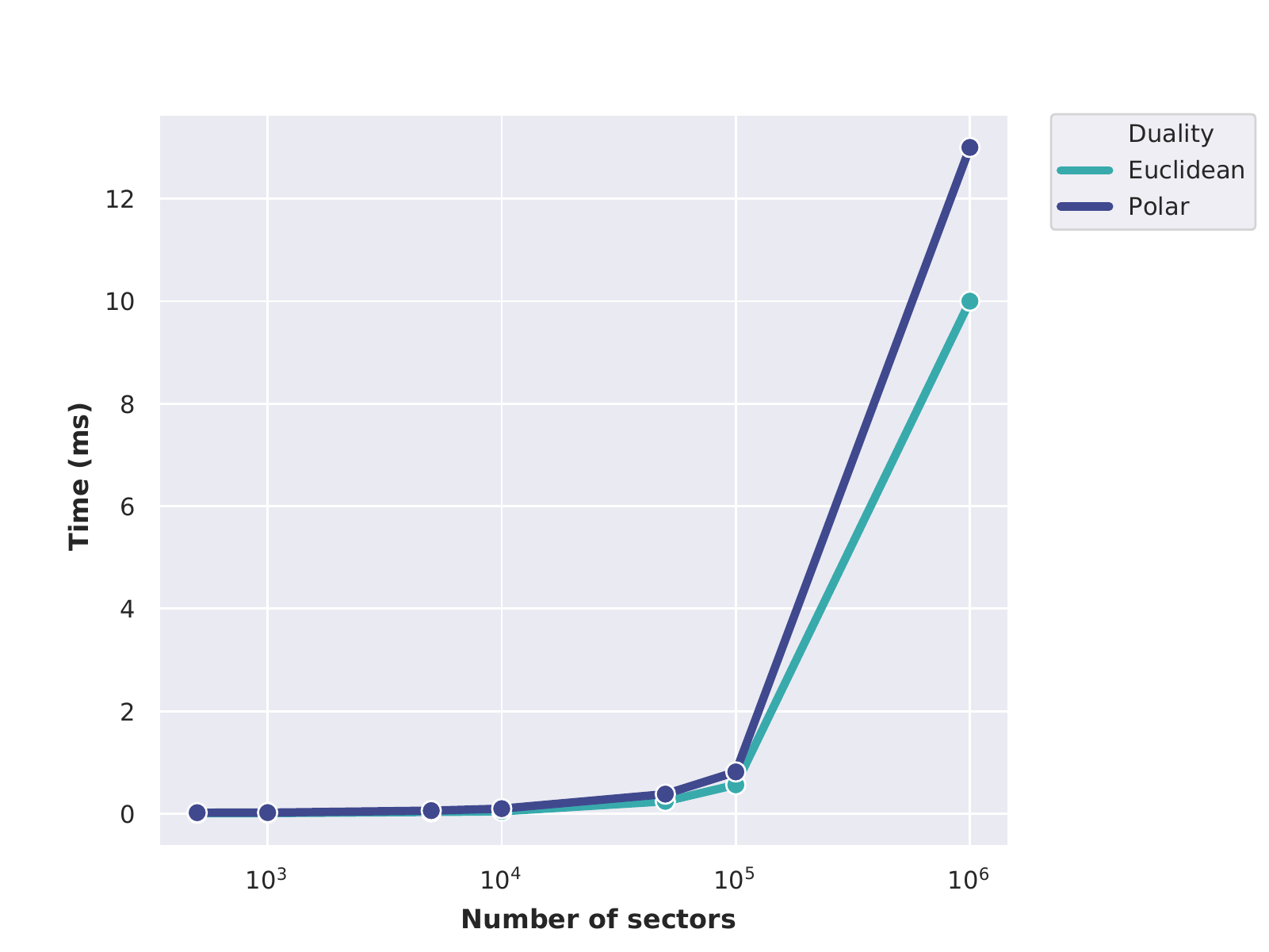}
  \caption{Average search time (in milliseconds, logarithmic scale) as
    a function of the number of setors, for the affine or the polar
    transforms. They both display the same behavior, with minor
    differences in search time.}
  \label{fig:compare_dualities}
\end{figure}

\paragraph{Choice for the dual transform}
In Section~\ref{sec:proposedmethod}, the dual R-tree is built from
either the affine or the polar transform. R-trees built from either
transform yield a similar average search time regardless of the number
of angular sectors (see Fig.~\ref{fig:compare_dualities}). For
example, for one million angular sectors, the average search time is
13 milliseconds (ms) for the R-tree built from the polar transform
versus 10 ms for the affine transform. However, two data structures
need to be maintained when the affine dual transform is used. The
effective cost in memory might be prohibitive as the number of sectors
increases. The polar dual transform only maintains one spatial data
structure so, in the remainder of the evaluation, we choose this dual
transform.

\paragraph{Datasets}
We used two types of datasets: synthetic data, and real data.

\begin{figure}[ht]
  \centering
  \includegraphics[width=0.7\textwidth]{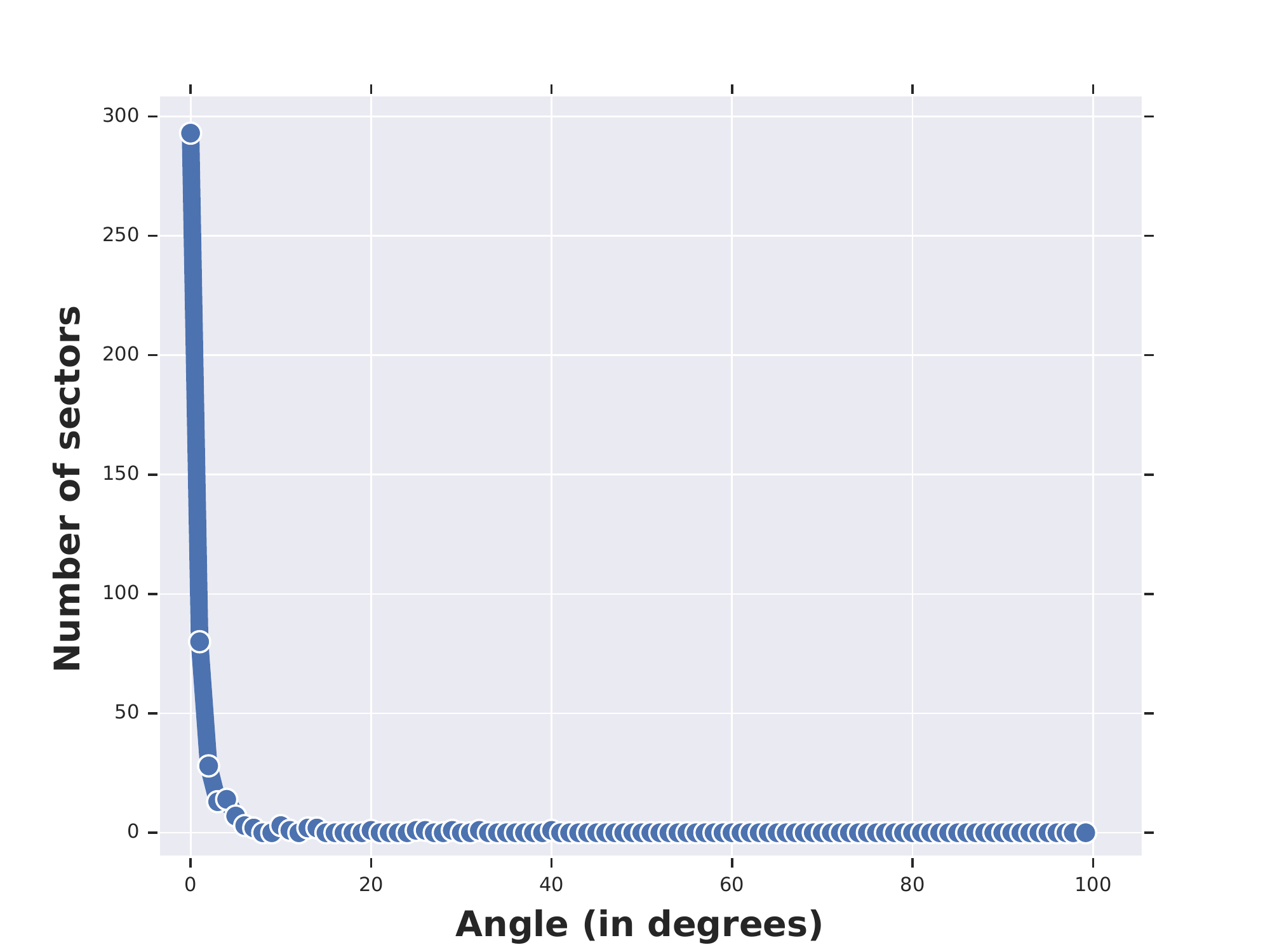}
  \caption{Angle distribution of angular sectors obtained from 500
    photographs.}
  \label{fig:angledistribution}
\end{figure}

The synthetic data have been generated by fitting a model on the
distribution of real data. The fields of view of 500 photographs,
taken in an urban environment, were used to compute their isovists.
Free visibility areas in the isovists constitute the angular sectors
(see Section~\ref{sec:intro}). The angle distribution for the 500
photographs is shown in Fig.~\ref{fig:angledistribution}. Synthetic
angular sectors were generated with angle values chosen to fit this
distribution.

The real dataset was obtained from geo-tagged photographs from both
Flickr\footnote{\url{https://www.flickr.com/services/api/}} and the
open building footprint dataset from the city of
Chicago\footnote{\url{https://data.cityofchicago.org/}}. The
photographs contain both position and direction information from which
the isovists were extracted. The \textbf{full} dataset consists in the
free visibility areas of the isovists (i.e. angular sectors). This
dataset contains photographs located outside the building footprint
area, in the country close to Chicago, typically. Thus, the associated
angular sectors effectively correspond to the full field of view of
the photograph. We generated an \textbf{urban} dataset by removing
these angular sectors from the full dataset. From the $\num{9.0e4}$
photographs polled from Flickr, the full dataset contains about
$\num{1.5e5}$ free visibility angular sectors, and the urban dataset
about $\num{1.2e5}$ elements.

In the following evaluation, the angular sectors need to be bounded to
be inserted inside the regular R-tree. They are represented as
polygons with a size of $\num{e8}$ meters, which is deemed
infinitely long at country scale.

\paragraph{Metrics and baseline methods}

We evaluate the search procedure on the dual R-tree, as the number of
sectors increases. Its efficiency is assessed by several metrics: the
initialization time (i.e. the time for the data structure to be fully
built), the average search time, the length of the returned list (in
number of elements), and the coverage and overlap measures inside the
spatial data structure (see Section~\ref{sec:previousworks}). The
coverage and overlap measures allow to quantify the amount of empty
space inside the R-tree, which is directly linked to the efficiency of
the search procedure. Let $r$ be a rectangle inside the R-tree, $C$
the child rectangles $r$ encloses,
$I_C = \{ c_i \cap c_j, \forall(c_i, c_j) \in C~s.t.~j > i\}$ the
total intersection between child rectangles, and $\text{Area}(s)$ the
area of an arbitrary shape $s$.

The coverage for $r$ is given by:

\[
  \text{coverage}(r) = \dfrac{\text{Area}(C)}{A(r)}
\]

The overlap inside $r$ is given by:

\[
  \text{overlap}(r) = \dfrac{\text{Area}(I_C)}{A(r)}
\]

Joint values of one for the coverage and 0 for the overlap means there
is no empty space inside a rectangle.

The point queries are made randomly in a region of 5x5 kilometers
around the average position of the angular sectors. We compare our
method to an \textbf{exhaustive search} (linear time) and to the
search in a \textbf{regular R-tree} from the equivalent angular
sectors. While the dual R-tree is built from bi-angular sectors, the
search by the baseline methods is made on mono-angular sectors, as
defined in Section~\ref{sec:intro}. This allows to restrict the
coverage and overlap for the regular R-tree, and is more in tune with
the initial query. The branching factor for the R-tree is fixed at 7
throughout the experiments, as it showed a good compromise between
number of access and size of the array resulting from the search
procedure (results not shown).

\subsection{Experimental results}

\subsubsection{Synthetic data}
There are no major discrepancies in the initialization time between the dual R-tree and regular R-tree (about 100ms on average).

\begin{figure}[ht]
  \begin{subfigure}[t]{0.5\linewidth}
    \centering
    \includegraphics[width=0.99\linewidth]{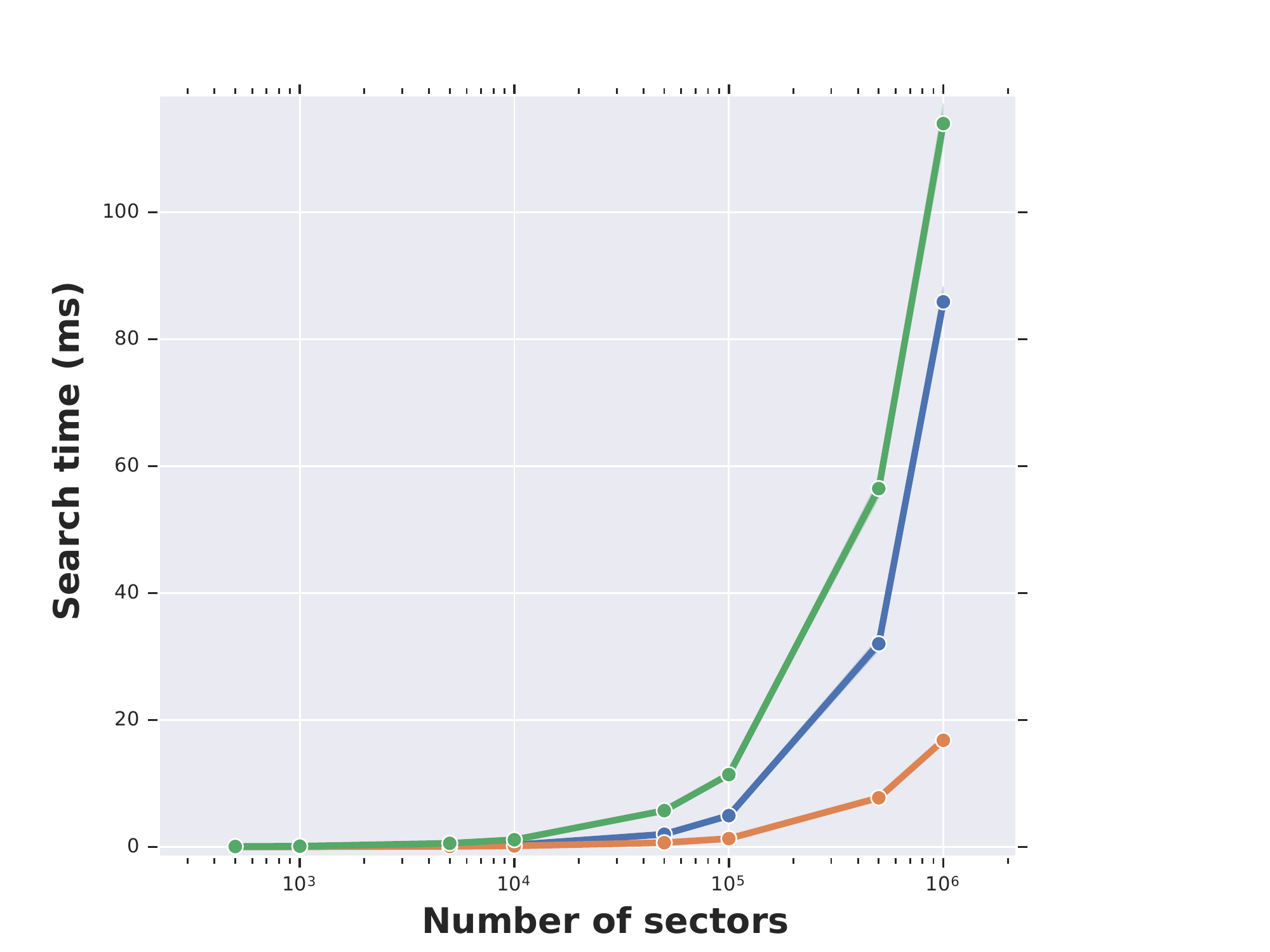}
    \caption{}
    \label{subfig:searchTime_time}
  \end{subfigure}%
  \begin{subfigure}[t]{0.5\linewidth}
    \centering
    \includegraphics[width=0.99\linewidth]{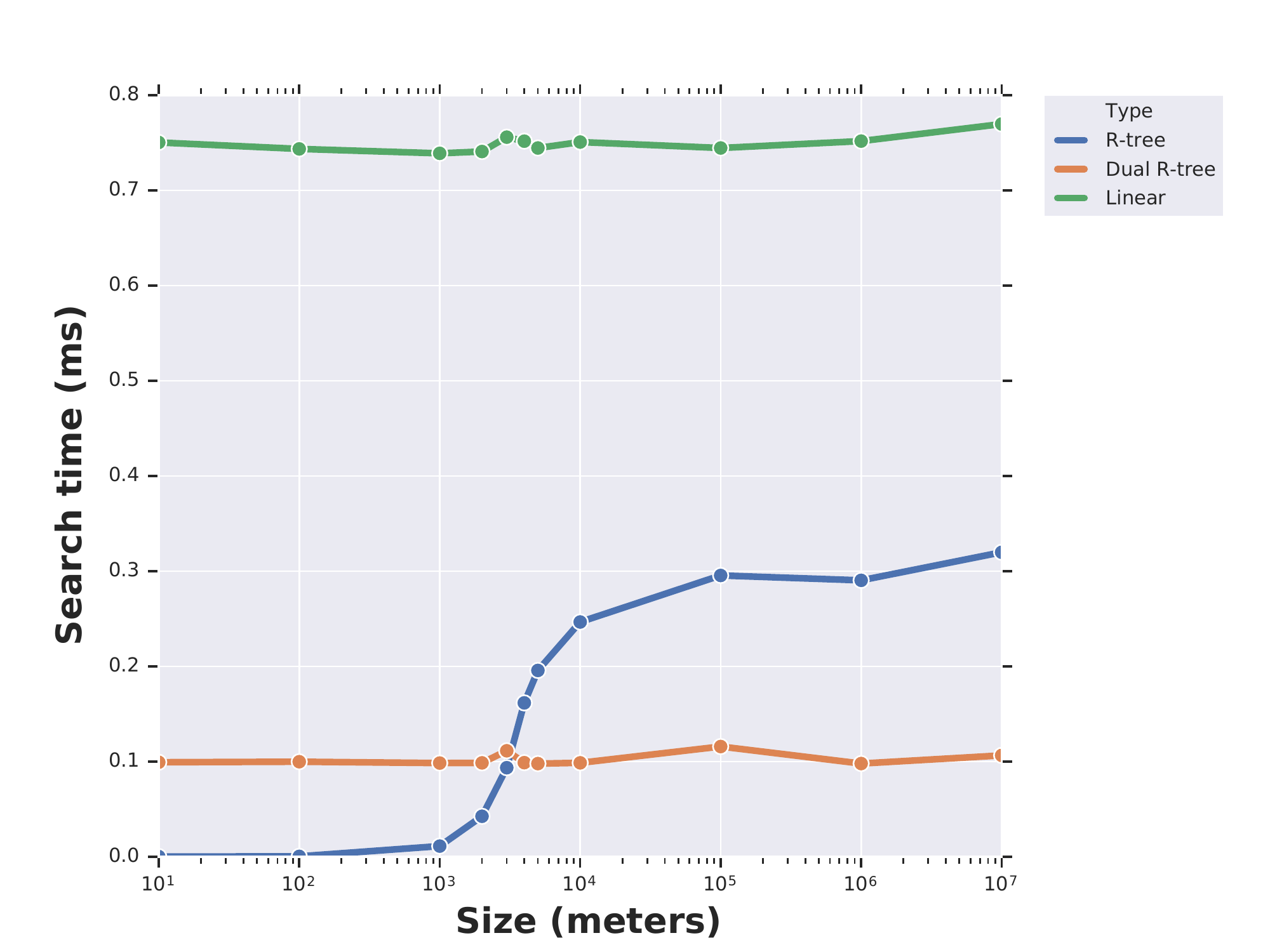}
    \caption{}
    \label{subfig:searchTime_radius}
  \end{subfigure}%

  \caption{Average search time (in milliseconds) as a function of (a)
    the number of sectors (logarithmic scale) and (b) the size (in
    meters, logarithmic scale) for the R-tree built from our method
    (Dual R-tree, in orange), the R-tree built from primal angular
    sectors (regular R-tree, in blue) and the exhaustive search
    (Linear, in green). (a) For one million angular sectors, the
    search procedure from the dual R-tree is five times faster than
    the regular R-tree. (b) Our method is faster for angular sectors
    which length is greater than 400 meters.}
  \label{fig:searchTime}
\end{figure}

The profile for the average search time on the synthetic data is shown
in Fig.~\ref{fig:searchTime}. For one million angular sectors, it is 5
times lower than the search time in the regular R-tree, and 6.5 times
lower than the search time for the exhaustive algorithm. However, the
profile shows our method does not achieve logarithmic query time.
Indeed, the search time is proportional to the number of sectors (see
Fig.~\ref{subfig:searchTime_time}). This is because angular sectors
are infinitely long geometrical shapes, so they permanently fill a
portion of the 2D space. For a large number $n$ of angular sectors of
angle $\alpha$ (in degrees), the average number of sectors going
through a random point is given by $\frac{n \times \alpha}{360}$. This
means the query time is proportional to $n$ and dependent on $\alpha$.
However, from our distribution, most angular sectors have a very small
angle: 65\% of sectors have an angle lower than one degree (see
Fig.~\ref{fig:angledistribution}) which makes our method faster than
the baseline methods.

We analyzed the search time as a function of the size (i.e. the radius) of the angular
sectors (see Fig.~\ref{subfig:searchTime_radius}). Our method is not
impacted by the size, but the regular R-tree is. When the size is
small, the angular sectors are reduced to small polygons for which the
regular R-tree is more efficient than our dual data structure. Our
method is more efficient for angular sectors of length greater than
500 meters. Thus, it is suited not only for infinitely long
geometrical shapes, but also for sufficiently large bounded ones.

\begin{figure}[ht]
  \centering
  \includegraphics[width=0.75\linewidth]{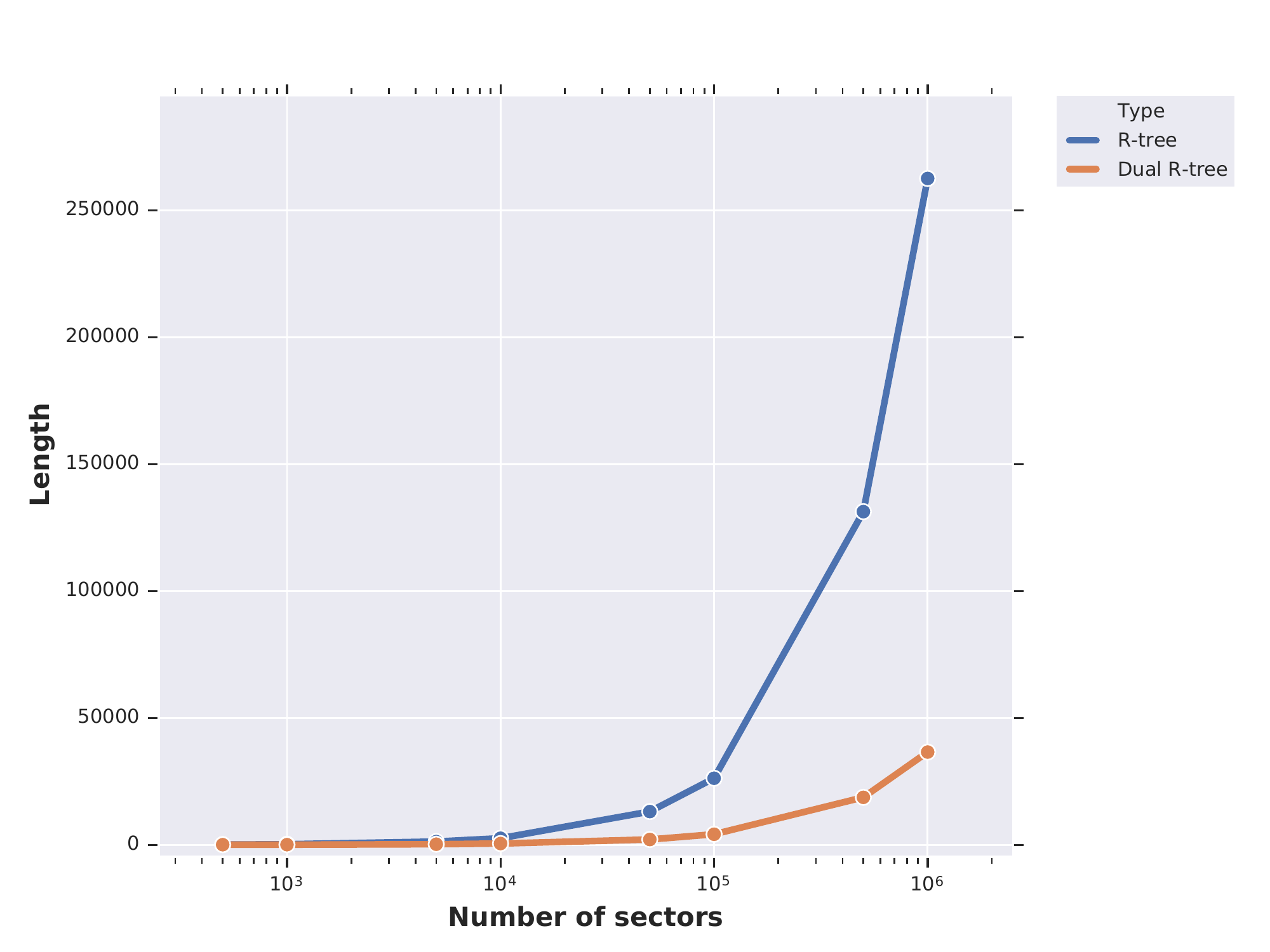}
  \caption{Length of the resulting array after the search, as a
    function of the number of sectors (logarithmic scale).}
  \label{fig:searchTime_size}
\end{figure}

The length of the returned array increases linearly with the number of
sectors for both the R-tree and dual R-tree (see
Fig.~\ref{fig:searchTime_size}). The returned array is 6.5 times
smaller with our method than with the regular R-tree. For one million
sectors, the number of elements is about $\num{4e4}$ for our method,
compared to $\num{2.6e5}$ for the regular R-tree. In order to remove the false positives
inside the array, it is necessary to process the resulting arrays
further. For the dual R-tree, this step is especially important in order to prune the
irrelevant symmetric parts of the bi-angular sectors (see
Section~\ref{sec:dual}).

\begin{figure}[ht]
  
  \begin{subfigure}[t]{0.5\linewidth}
    \centering
    \includegraphics[width=0.99\linewidth]{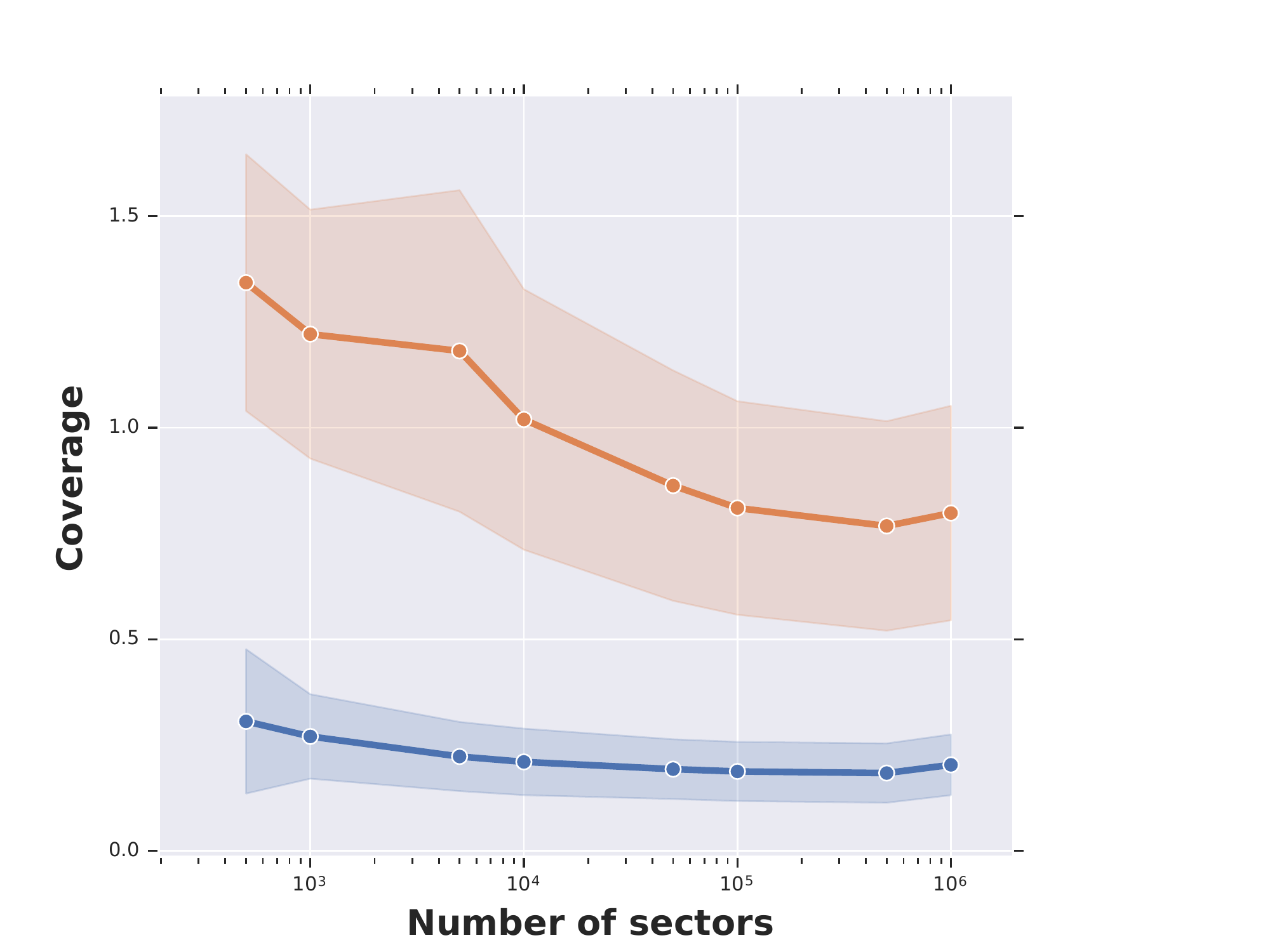}
    \caption{}
    \label{subfig:coverage}
  \end{subfigure}%
  \begin{subfigure}[t]{0.5\linewidth}
    \centering
    \includegraphics[width=0.99\linewidth]{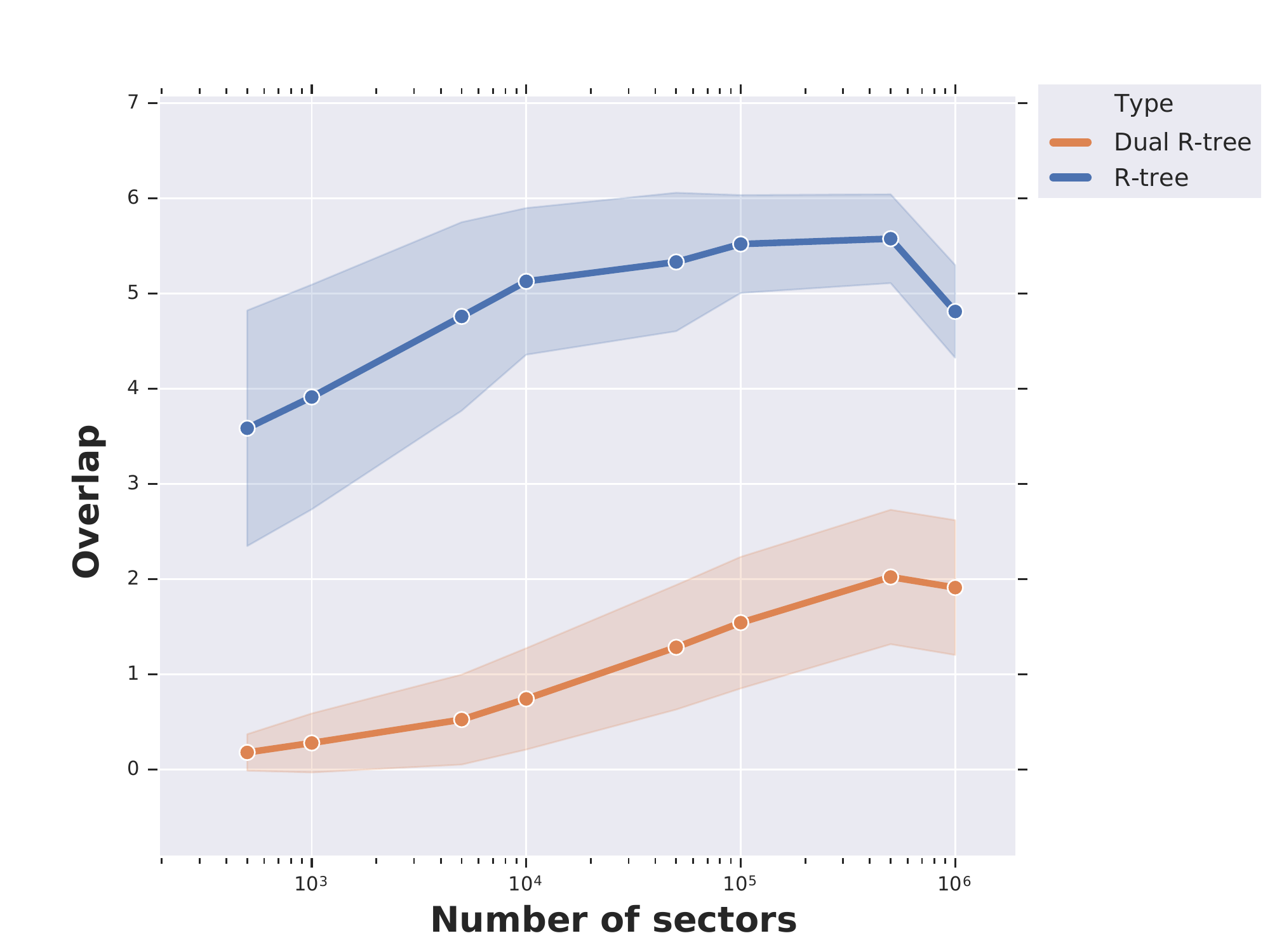}
    \caption{}
    \label{subfig:overlap}
  \end{subfigure}%

  \caption{Average (a) coverage and (b) overlap measures as a function
    of the number of sectors (logarithmic scale), for the R-tree built
    from our method, with dual angular sectors (Dual R-tree, in
    orange), and the R-tree built from primal angular sectors (regular
    R-tree, in blue). (a) The dual R-tree has coverage values closer
    to one compared to the regular R-tree. (b) The overlap values for
    the dual R-tree are lower than those of the regular R-tree. For
    $\num{e6}$ sectors, there is a non-proportional increase in the
    number of available rectangles in the lower levels of the
    hierarchy. Thus, the overlap decreases for both trees. }
  \label{fig:coverage}
\end{figure}

Results for the average coverage and overlap measures for all nodes in
the tree are shown in Figure~\ref{fig:coverage}. The coverage values
for the dual R-tree are closer to one than those of the regular
R-tree. This is combined with low overlap values for the dual R-tree,
which means there is less empty space inside the tree. It allows for
faster elimination of negative results during the search procedure.
This translates in an efficient data structure allowing to quickly
search through angular sectors.

\subsubsection{Real data}

The results for the real dataset, for photographs from the city of
Chicago is shown in Table~\ref{tab:real}.

\begin{table}[ht]
  \centering
  \caption{Results obtained by our method compared to the regular
    R-tree and an exhaustive search on angular sectors from 
    photographs of Chicago (Flickr data), for the full dataset ($\num{1.5e5}$
    angular sectors) and the urban dataset ($\num{1.2e5}$ angular
    sectors). Our method provides lower search times for both
    datasets, and better values for both the coverage and the
    overlap.}
  \label{tab:real}
  \begin{tabular}{llllll}
    \hline
    &             & \begin{tabular}[c]{@{}l@{}}Build\\ time\\(ms)\end{tabular} & \begin{tabular}[c]{@{}l@{}}Average \\ search time\\(ms)\end{tabular} & Coverage & Overlap \\ \hline
    \multirow{3}{*}{Full}  & Linear      &    N/A                                                         &    9.41                                                            &    N/A      &  N/A       \\
    & R-tree      &    175                                                           &                               6.20                                 &  0.26        &   8.79      \\
    & Dual R-tree &   199                                                            &                                3.43                                &  0.87        &   2.02      \\ \hline
    \multirow{3}{*}{Urban} & Linear      &    N/A                                                           &          6.84                                                      &  N/A        &  N/A       \\
    & R-tree      &     130                                                          &              3.03                                                  &  0.30        &   8.27      \\
    & Dual R-tree &     165                                                       &                   0.776                                             &  1.08        &   1.59      \\ \hline
  \end{tabular}
\end{table}

Our method is more efficient for both datasets, compared to baseline
methods. The average search time is lower, the coverage is close to
one, and the overlap is lower than the regular R-tree. Nevertheless,
the efficiency of the dual R-tree is not as obvious for the full
dataset, as the search time is half of the search time on a regular
R-tree. This dataset contains a large number of angular sectors with
high angle values so it does not fulfill the hypotheses made on angle
values, and is, as such, not well-suited to our method. The results are
provided for the sake of completeness, and prove our method is better
suited for angular sectors with low angle values, typically found from
photographs in a city, rather than in the country.

The results obtained on the urban dataset are in line with those
obtained on the synthetic dataset. The average search time is four
times lower than that of the regular R-tree, and the better values for
the coverage and overlap mean there is less empty space inside the
tree.

\section{Conclusion}

In this article, we have introduced a new method to search through
angular sectors. The infinitely long angular sectors are converted to finite
coordinates by dual transforms. Then, a spatial data structure is
built from these coordinates. In Section~\ref{sec:evaluation}, we explained why the search procedure
cannot achieve $\mathcal{O}(\log n)$ time. However, the results show
our method provide faster results than existing spatial data
structures when searching through a large number of angular sectors.
Moreover, our method is not dependent on the size of the angular
sectors, and is suited to sufficiently-large bounded polygons. It can
be used to search efficiently through free-visibility areas of
isovists in a city.

The amount of empty space inside the dual R-tree could be reduced by
taking into account the shape of the dual angular sector. For instance, our method could be improved by transposing the OR-tree (see
Section \ref{sec:previousworks}) to dual coordinates.

Our method is suited for 2D angular sectors. Recently, 3D city models have
become more and more prominent, and 3D isovists can be extracted from
them. We plan to extend our method to 3D, by transforming the 3D
angular sectors with an adapted dual transform, such as the one from
\cite{Borrmann10}.

\section*{Acknowledgements}

This work was part of the Optimum project (Observatoire photographique
du territoire : images des mondes urbains en mutation) led dy
Dani\`{e}le M\'{e}aux and with the fruitful participation of Guillaume
Bonnel. This project was supported by the LABEX IMU (ANR-10-LABX-0088)
of Universit\'{e} de Lyon, within the program ``Investissements
d'avenir'' (ANR-11-IDEX-0007) operated by the French National Research
Agency (ANR).

\bibliography{dualrtree}

\begin{thebibliography}{}

\bibitem[Ay et~al., 2008]{Ay08}
Ay, S.~A., Zimmermann, R., and Kim, S.~H. (2008).
\newblock Viewable scene modeling for geospatial video search.
\newblock In {\em Proceeding of the 16th {ACM} international conference on
  Multimedia - {MM} {\textquotesingle}08}. {ACM} Press.

\bibitem[Benedikt, 1979]{Benedikt79}
Benedikt, M. (1979).
\newblock To take hold of space: Isovists and isovist fields.
\newblock {\em Environment and Planning B: Planning and Design}, 6:47--65.

\bibitem[Bentley, 1975]{Bentley75}
Bentley, J.~L. (1975).
\newblock Multidimensional binary search trees used for associative searching.
\newblock {\em Commun. ACM}, 18(9):509--517.

\bibitem[Borrmann et~al., 2010]{Borrmann10}
Borrmann, D., Elseberg, J., Lingemann, K., and Nüchter, A. (2010).
\newblock A data structure for the 3{D} {H}ough {T}ransform for plane
  detection.
\newblock {\em IFAC Proceedings Volumes}, 43(16):49 -- 54.
\newblock 7th IFAC Symposium on Intelligent Autonomous Vehicles.

\bibitem[Comer, 1979]{Comer79}
Comer, D. (1979).
\newblock Ubiquitous {B}-{T}ree.
\newblock {\em ACM Comput. Surv.}, 11(2):121--137.

\bibitem[Finkel and Bentley, 1974]{Finkel74}
Finkel, R.~A. and Bentley, J.~L. (1974).
\newblock Quad trees a data structure for retrieval on composite keys.
\newblock {\em Acta Informatica}, 4(1):1--9.

\bibitem[Guttman, 1984]{Guttman84}
Guttman, A. (1984).
\newblock R-trees: A dynamic index structure for spatial searching.
\newblock {\em SIGMOD Rec.}, 14(2):47--57.

\bibitem[Lawson-Peebles, 1988]{Lawson88}
Lawson-Peebles, R. (1988).
\newblock John brinckerhoff jackson, discovering the vernacular landscape (new
  haven {\&} london: Yale university press, 1984, {\textsterling}14.95) pp.
  166. {ISBN} 0 300 03138 6. - richard p. horwitz, the strip: An american
  place. photographs by karin e. becker. (lincoln {\&} london: University of
  nebraska press, 1985, {\textdollar}14.95). pp. 188. {ISBN} 8032 7228 6.
\newblock {\em Journal of American Studies}, 22(1):149--151.

\bibitem[Lu and Shahabi, 2017]{Lu17}
Lu, Y. and Shahabi, C. (2017).
\newblock Efficient indexing and querying of geo-tagged aerial videos.
\newblock In {\em Proceedings of the 25th {ACM} {SIGSPATIAL} International
  Conference on Advances in Geographic Information Systems -
  {SIGSPATIAL}{\textquotesingle}17}. {ACM} Press.

\bibitem[Lu et~al., 2014]{Lu14}
Lu, Y., Shahabi, C., and Kim, S.~H. (2014).
\newblock An efficient index structure for large-scale geo-tagged video
  databases.
\newblock In {\em Proceedings of the 22nd {ACM} {SIGSPATIAL} International
  Conference on Advances in Geographic Information Systems - {SIGSPATIAL}
  {\textquotesingle}14}. {ACM} Press.

\bibitem[Lu et~al., 2016]{Lu16}
Lu, Y., Shahabi, C., and Kim, S.~H. (2016).
\newblock Efficient indexing and retrieval of large-scale geo-tagged video
  databases.
\newblock {\em {GeoInformatica}}, 20(4):829--857.

\bibitem[Ma et~al., 2013]{Ma13}
Ma, H., Ay, S.~A., Zimmermann, R., and Kim, S.~H. (2013).
\newblock Large-scale geo-tagged video indexing and queries.
\newblock {\em {GeoInformatica}}, 18(4):671--697.

\bibitem[Nievergelt et~al., 1984]{Nievergelt84}
Nievergelt, J., Hinterberger, H., and Sevcik, K.~C. (1984).
\newblock The grid file: An adaptable, symmetric multikey file structure.
\newblock {\em {ACM} Transactions on Database Systems}, 9(1):38--71.

\bibitem[Okabe et~al., 2000]{Wiley00}
Okabe, A., Boots, B., Sugihara, K., Chiu, S.~N., and Kendall, D.~G., editors
  (2000).
\newblock {\em Spatial Tessellations}.
\newblock John Wiley {\&} Sons, Inc.

\bibitem[Poncelet, 1865]{Poncelet1865}
Poncelet, J. (1865).
\newblock {\em Trait{\'e} des propri{\'e}t{\'e}s projectives des figures:
  ouvrage utile {\`a} ceux qui s'occupent des applications de la
  g{\'e}om{\'e}trie descriptive et d'op{\'e}rations g{\'e}om{\'e}triques sur le
  terrain}.
\newblock Number~1. Gauthier-Villars.

\bibitem[Radon, 1917]{Radon1917}
Radon, J. (1917).
\newblock On the determination of functions from their integral values along
  certain manifolds.
\newblock {\em IEEE Transactions on Medical Imaging}, 5(4):170--176.

\bibitem[Sellis et~al., 1987]{Sellis87}
Sellis, T.~K., Roussopoulos, N., and Faloutsos, C. (1987).
\newblock The {R}+-{T}ree: A dynamic index for multi-dimensional objects.
\newblock In {\em Proceedings of the 13th International Conference on Very
  Large Data Bases}, VLDB '87, pages 507--518, San Francisco, CA, USA. Morgan
  Kaufmann Publishers Inc.

\bibitem[Suleiman et~al., 2013]{Suleiman13}
Suleiman, W., Joliveau, T., and Favier, E. (2013).
\newblock {\em A New Algorithm for 3{D} Isovists}, pages 157--173.
\newblock Springer Berlin Heidelberg, Berlin, Heidelberg.

\bibitem[Turner et~al., 2001]{Turner01}
Turner, A., Doxa, M., O'Sullivan, D., and Penn, A. (2001).
\newblock From isovists to visibility graphs: A methodology for the analysis of
  architectural space.
\newblock {\em Environment and Planning B: Planning and Design},
  28(1):103--121.

\end{thebibliography}

\end{document}